\documentclass[english]{article}
\usepackage[T1]{fontenc}
\usepackage{color}
\usepackage{graphicx}
\usepackage{a4wide}
\usepackage{times}
\usepackage{float}
\usepackage{babel}

\makeatletter

\newcommand{\noun}[1]{\textsc{#1}}

\providecommand{\tabularnewline}{\\}

\input{epsfig.sty}
\def\fnum@table{\tablename~{\bf\thetable}}
\def\fnum@figure{\figurename~{\bf\thefigure}}
\def\tablename{\footnotesize{\bf Table}}
\def\figurename{\footnotesize{\bf Figure}}

\def\be{\begin{equation}}
\def\ee{\end{equation}}

\makeatother

\begin{document}

\title{
\textbf{
One-dimensional Hybrid Approach to Extensive Air Shower Simulation
}\\[1cm]
}

\author{
T. Bergmann$^{1}$, R. Engel$^{1}$, D. Heck$^{1}$, N. N. Kalmykov$^{2}$,\\
S. Ostapchenko\footnote{corresponding author, e-mail: serguei@ik.fzk.de} $^{1,2}$,
T. Pierog$^{1}$, T. Thouw$^{1}$,  K. Werner$^{3}$
\\
\textit{$^{1}$}\textit{\small Forschungszentrum Karlsruhe, Institut
f\"ur Kernphysik, 76021 Karlsruhe, Germany}\\
 \textit{$^{2}$}\textit{\small D.V. Skobeltsyn Institute of Nuclear
Physics, Moscow State University, 119992 Moscow, Russia}\\
\textit{$^{3}$}\textit{\small SUBATECH, Universit\'e de Nantes -- IN2P3/CNRS
-- Ecole des Mines, Nantes, France}\\
} 

\maketitle

\begin{center}
\textbf{\large Abstract}
\end{center}
{\large \par}

An efficient scheme for one-dimensional extensive air shower simulation
and its implementation in the program {\sc conex} are presented. 
Explicit Monte Carlo simulation
of the high-energy part of hadronic and electromagnetic cascades
in the atmosphere is combined with a numeric solution of 
cascade equations for smaller energy sub-showers to obtain accurate 
shower predictions. The developed scheme allows us to calculate not only
observables related to the number of particles (shower size)
but also ionization energy deposit profiles which are needed for the
interpretation of data of experiments employing the fluorescence light
technique. We discuss in detail the basic algorithms developed and illustrate
the power of the method. It is shown that Monte Carlo, numerical,
and hybrid air shower calculations give consistent results which agree
very well with those obtained within the \textsc{corsika} program. 


\section{Introduction\label{intro.sec} }

Monte Carlo (MC) simulation of extensive air showers (EAS) is the
most common method to calculate detailed theoretical predictions 
needed for interpreting experimental data of air shower arrays
or fluorescence light detectors. However, for primary particles of very
high energy, straight-forward MC simulation is not a viable option 
because of the unreasonably
large computing time required. The situation can be improved by applying
some weighted sampling algorithms, like the so-called ``thinning''
method \cite{Hillas81a}, i.e.\ treating explicitly only a small portion
of all shower particles and assigning each particle a corresponding
weight factor. Although this approach allows the reduction of EAS
calculation times to practically affordable values, it comes soon
to its limits. The summation of particle contributions with very large
weights creates significant artificial fluctuations for EAS observables
of interest \cite{Kobal99a,Risse01a,Kobal:2001jx}. Imposing maximum weight limitations to
ensure high simulation quality \cite{Kobal:2001jx}, on the other hand, 
prevents one from using less detailed sampling and correspondingly
from further speeding-up the calculation process. A possible alternative
procedure is to describe EAS development numerically, based on the
solution of the corresponding cascade equations 
\cite{Dedenko65a,Hillas65a,Bossard:2000jh}.
Combining this with an explicit MC simulation of the most high-energy
part of an air shower allows one to obtain accurate results
both for average EAS characteristics and for their fluctuations
\cite{KaM13}. 

In this article we describe a new EAS simulation program of such a
hybrid type, called \noun{conex}. In \noun{conex} the MC treatment of
above-threshold particle cascading is realized in the standard way
and does not differ significantly from the implementation in e.g.\
\textsc{corsika} \cite{Heck98a}. On the other hand, the numerical description
of lower energy sub-cascades is based on the solution of hadronic
cascade equations, using an updated algorithm of Ref.~\cite{Bossard:2000jh},
and a newly developed procedure for the solution of electro-magnetic
(e/m) cascade equations. The corresponding algorithms are characterized
by high efficiency and good accuracy even if a 
comparatively crude binning with respect to particle energy
and depth position is used.
Furthermore, by accounting for neutrino production in addition to the typically
considered particles, \noun{conex} can also be used for the calculation
of ionization energy deposit profiles.

The numeric solution of cascade equations can be replaced, in principle,
by a pre-tabulation of the characteristics of secondary sub-cascades,
obtained via an iterative MC procedure 
\cite{Stanev:1994pf,Gaisser97a,Alvarez-Muniz:2002ne}. An
example of a combined approach, extended to a three-dimensional EAS
simulation, is described in Ref. \cite{Drescher:2002cr}, where hadronic sub-cascades
are treated using the method of Ref. \cite{Bossard:2000jh} and e/m sub-cascades
are tabulated using the \textsc{egs4} MC program \cite{egs4}. 

The approach presented in the current work does not require any pre-tabulation
of particle cascades and is characterized by high efficiency and large
flexibility. It can be applied to various initial conditions, i.e.\
a wide range of energies and angles of incidence of a primary particle,
including the case of upward-going showers, as well as arbitrary parameterizations
of the atmosphere of the Earth. These features make \noun{conex}
ideally suited for applications related to the event-by-event based
analysis of EAS data, in particular, of fluorescence light based measurements.

This is the first paper in a series in which we will investigate various 
features of EAS and their relation to the characteristics of hadronic 
multiparticle production. In this work, we shall present the hybrid simulation 
scheme in detail, leaving the study of shower predictions to a forthcoming article. 

The outline of the paper is as follows. Section \ref{scheme} describes
the calculation scheme and its basic procedures. In Section \ref{eas-char}
we show some examples of calculated EAS characteristics and investigate
the accuracy and efficiency of the method comparing the hybrid approach
with pure MC or numerical procedures. The reliability of the predictions
is checked by a detailed comparison with \textsc{corsika} results. Finally, a
summary is given in Section \ref{sec:Summary} and both potential
applications of the program and the prospects for its further development
are discussed.


\section{Calculation scheme\label{scheme}}

\subsection{Physics overview}

The calculation scheme consists
of two main stages: an explicit MC simulation of the cascade for
particles
with energies above some chosen threshold $E_{\rm thr}$ (being a
free parameter of the scheme) and a solution of nuclear-electro-magnetic
cascade equations for sub-cascades of smaller energies. Both MC and
numerical parts are characterized by the same physics content, as
described below. 

In the hadronic cascade one follows the propagation, interaction
and decay (where applicable) of (anti-) nucleons, charged pions, charged
and neutral kaons; all other types of hadrons produced in interactions
and decays are assumed to decay immediately. Particle interactions
in the MC part are treated within a chosen high energy hadronic interaction
model (implemented are 
\textsc{nexus 3.97}~\cite{Drescher:2000ha,Pierog:2002gj,Werner:2003},
\textsc{qgsjet 01}~\cite{Kalmykov:1994ys,Kalmykov:1997te} and 
\textsc{ii} \cite{Ostapchenko:2004ss,Ostapchenko:2006vr,Ostapchenko:2005yj},
and \textsc{sibyll 2.1}~\cite{Engel:1992vf,Fletcher:1994bd,Engel99a}), 
decays are simulated using
the corresponding routines of the \textsc{nexus} model. The same models
are used to pre-calculate secondary particle spectra for later
use in the numerical treatment of hadronic cascade equations. Optionally,
below some energy $E_{\rm low}^{\rm had}\sim100$ GeV, 
\textsc{gheisha}~\cite{Fesefeldt85a} is employed as low-energy hadronic interaction model. 

The MC treatment of the e/m cascade is realized by means of the \textsc{egs4}
code \cite{egs4}, supplemented by an account of the Landau-Pomeranchuk-Migdal
(LPM) effect \cite{Migdal:1956tc,Konishi:1990ya,Heck:1998gr} for
ultra-high energy electrons (positrons) and photons. The simulation of
photonuclear interactions and muon pair production was added to the
\textsc{egs} package closely following the \textsc{corsika} implementation
 \cite{Heck98a}.
The system
of coupled e/m cascade equations is based on the same interaction
processes as implemented in the MC, using Bethe-Heitler cross sections
for the bremsstrahlung and pair production with energy-dependent 
correction factors according
to Storm and Israel \cite{Storm70a}, the Klein-Nishina formula for the
Compton process, and accounting for Moeller and Bhabha processes
as well as for positron-electron annihilation 
(see, for example \cite{egs4,Gaisser90a}).
Both LPM suppression and photo-effect are neglected in the e/m cascade
Eqs., as the latter are employed in the energy ranges where these 
processes are not important. Ionization losses of electrons
and positrons are described by the Bethe-Bloch formula with corrections
to account for the density effect \cite{Sternheimer:1983mb}.

High energy interactions of muons (bremsstrahlung, pair production
and muon-nuclear interaction
\cite{Bottai:2000en,Andreev:1997pf,Bezrukov:1981ci,Bilokon:1990nc,Lohmann:1985qg})
are taken into account in the MC part but are neglected in the cascade
Eqs.

In general, an individual shower is simulated as follows. One starts
with the primary particle of given energy, direction
and initial position in the atmosphere (by default, at 100 km above sea
level, if no special geometry is required, e.g.\ up-going showers).
The initial particle direction thus defines the position of the ``true''
shower axis, in the following referred to as the ``shower trajectory''.
For a hadron as primary particle, one simulates the hadronic cascade
explicitly, recording all secondary particles at a number of pre-chosen
depth levels and energy intervals, until all produced secondaries
have an energy lower than the threshold $E_{\rm thr}$. The levels are defined
with respect to the projected depth $X$, i.e.\ the slant depth
for the particle position projected to the initial shower axis (shower
trajectory), as described in more detail in the Appendix. 
All sub-threshold hadrons/muons and e/m particles
are filled into energy-depth tables
that form the ``source terms'' for the cascade equations.
In parallel, the above-threshold e/m particles are transferred to
\textsc{egs4} for simulating the e/m particle cascade
in a similar way, with all sub-threshold e/m particles
being added to the e/m source terms. 

In the next step the hadronic
cascade at energies below $E_{\rm thr}$ is calculated numerically
for the first depth level using the 
corresponding cascade equations and initial
conditions specified by the source terms. As the result, one obtains
discretized energy spectra of hadrons of different types at the next
depth level. All sub-threshold e/m particles produced at this stage
are added to the e/m source term. Then sub-threshold e/m cascades
are calculated by solving the corresponding e/m cascade equations
for the given initial conditions.  Hadrons due to photonuclear
interaction and pair-produced muons that are generated in the numerical
solution of the e/m cascade Eqs.\ are added to the hadronic source term
of the next slant depth level. 
This procedure is repeated for the following depth levels, each time
using the hadronic and e/m source terms of the previous level.

Ultra-high energy e/m particles can undergo geomagnetic pair production and
bremsstrahlung well above the atmosphere of the Earth
\cite{Erber:1966vv,Stanev:1996ux,Vankov:2002cb}. Therefore,
in case of the primary particle being a photon or an electron, the
simulation process starts with the calculation of possible interactions
with the geomagnetic field using the \noun{preshower} code \cite{Homola:2003ru}
and the above described procedure is applied to the secondary
particles.


\subsection{Hadronic cascade equations\label{sub:Hadronic-cascade-equations}}

The backbone of a hadron-initiated extensive air shower is the hadronic
cascade which develops via particle propagation, decay, and interaction with air nuclei
of both the initial particle and of produced secondary hadrons. The
corresponding integro-differential equations are given by \cite{Bossard:2000jh}
(see also \cite{Drescher:2002cr})
\begin{eqnarray}
\frac{\partial\left.h_{a}(E,X)\right|_{T}}{{\partial
X}}&=&-\frac{\left.h_{a}(E,X)\right|_{T}}{\lambda_{a}(E)}-\left.h_{a}(E,X)
\right|_{T}\frac{\left|\frac{dL}{dX}\right|_{T}}{\tau_{a}(E)\,
c}
+\frac{\partial}{\partial E}\left(\beta_{a}^{{\rm
ion}}(E) \left.h_{a}(E,X)\right|_{T}\right)\nonumber \\ 
&  & +\sum_{d}\int_{E}^{E_{\max}}\!
dE'\;\left.h_{d}(E',X)\right|_{T}\;\left[\frac{W_{d\rightarrow
a}(E',E)}{\lambda_{d}(E')}+D_{d\rightarrow
a}(E',E)\frac{\left|\frac{dL}{dX}\right|_{T}}{\tau_{d}(E')\,
c}\right]\nonumber \\
 &  & +\left.S_{a}^{{\rm had}}(E,X)\right|_{T},
\label{sys1-had}
\end{eqnarray}
where $\left.h_{a}(E,X)\right|_{T}$ are the differential energy spectra
of hadrons of type $a$ with energy $E$ at depth position $X$ along
a given straight line trajectory $T$ (in the following the $T$-symbol
will be omitted), $\beta_{a}^{{\rm ion}}(E)=-dE_{a}/dX$
is the ionization energy loss of particle $a$ per depth unit. A muon
is treated like a hadron, but without interaction term. 

The first term in the r.h.s.\ of Eq.~(\ref{sys1-had}) represents the
decrease of hadron number due to interactions with air nuclei
\begin{equation}
\frac{dh_{a}}{dX}=-\frac{h_{a}}{\lambda_{a}},
\end{equation}
with the corresponding mean free path 
$\lambda_{a}=m_{\mathrm{air}}/\sigma_{\mathrm{inel}}^{a-{\rm air}}$,
where $m_{{\rm air}}$ is the average mass of air molecules and $\sigma_{\mathrm{inel}}^{a-{\rm air}}$
is the hadron $a$ - air nucleus inelastic cross section.

The second term describes particle decay, with the decay rate on a
path $dL$ being 
\begin{equation}
dh_{a}=-h_{a}\:\frac{dL}{\tau_{a}\, c}\label{dhdt},
\end{equation}
where $\tau_{a}$ is the life time of hadron $a$ in the lab.\ system,
related to the proper life time $\tau_{a}^{(0)}$ by $\tau_{a}=\tau_{a}^{(0)}\, E/m_{a}$,
with $m_{a}$ being the hadron mass and $c$ the velocity of light.
From the definition of slant depth (\ref{depth}) follows 
\begin{equation}
\left|\frac{dL}{dX}\right|=\frac{1}{\rho_{{\rm air}}(X)}\label{dldx}.
\end{equation}

The third term in Eq.~(\ref{sys1-had}) takes into account 
particle ionization energy losses and 
the integral term in Eq. (\ref{sys1-had}) represents the production of
particles of type $a$ in interactions and decays of higher energy
parents of type $d$, with $W_{d\rightarrow a}(E',E)$, $D_{d\rightarrow a}(E',E)$
being the corresponding inclusive spectra of secondaries. 

Finally, the so-called source term $S_{a}^{{\rm had}}(E,X)$ defines
the initial conditions and is determined during the MC simulation
of above-threshold particle cascading. It consists of contributions
of all sub-threshold hadrons produced at that stage
\begin{equation}
S_{a}^{{\rm had}}(E,X)=S_{a}^{{\rm MC\rightarrow
had}}(E,X)=\sum_{i=1}^{N_{{\rm source}}^{{\rm
had}}}\delta_{d_{i}}^{a}\,\delta(E-E_{i})\,\delta(X-X_{i}),\label{source1}
\end{equation} 
with $d_{i}$, $E_{i}$, $X_{i}$ being type, energy, and
depth position of the source particles. 

The numerical method of solving the hadronic cascade equations is similar to 
the approach of \cite{Bossard:2000jh} and is 
summarized in Appendix \ref{sub:Numerical-had}.


\subsection{Electro-magnetic cascade equations}

The e/m cascade development can be described by the following system
of integro-differential equations (see, for example, \cite{Gaisser90a})
\begin{eqnarray} 
\frac{\partial l_{e^{-}}\!(E,X)}{{\partial
X}}&=&-\sigma_{e^{-}}\!(E)\: l_{e^{-}}\!(E,X)
+\frac{\partial}{\partial E}\left( \beta_{e^{-}}^{{\rm ion}}(E) l_{e^{-}}(E,X)\right)
\label{sys1-el}
\nonumber \\ 
&  & +\int_{E}^{E_{\max}}\!
dE'\,\left[l_{e^{-}}\!(E',X)\, W_{e^{-}\rightarrow
e^{-}}\!(E',E)+l_{e^{+}}\!(E',X)\right.
\nonumber \\ 
&  & \times\left.\,
W_{e^{+}\rightarrow e^{-}}\!(E',E)+l_{\gamma}\!(E',X)\,
W_{\gamma\rightarrow e^{-}}\!(E',E)\right]+S_{e^{-}}^{{\rm
e/m}}\!(E,X)
\\ 
\nonumber \\ 
\frac{\partial
l_{e^{+}}\!(E,X)}{{\partial X}}&=&-\sigma_{e^{+}}\!(E)\:
l_{e^{+}}\!(E,X)
+\frac{\partial}{\partial E}\left(\beta_{e^{+}}^{{\rm ion}}(E)
l_{e^{+}}(E,X)\right)
\nonumber \\ 
&  & +\int_{E}^{E_{\max}}\!
dE'\,\left[l_{e^{+}}\!(E',X)\, W_{e^{+}\rightarrow
e^{+}}\!(E',E)\right.
\nonumber \\ 
&  & +\left.l_{\gamma}\!(E',X)\,
W_{\gamma\rightarrow e^{+}}\!(E',E)\right]+S_{e^{+}}^{{\rm
e/m}}\!(E,X)
\\ 
\frac{\partial
l_{\gamma}\!(E,X)}{{\partial X}}&=&-\sigma_{\gamma}\!(E)\:
l_{\gamma}\!(E,X)+\int_{E}^{E_{\max}}\! dE'\label{sys1-gam}
\nonumber \\ 
&  & \times\left[l_{\gamma}\!(E',X)W_{\gamma\rightarrow\gamma}
\!(E',E)+l_{e^{-}}\!(E',X)\,
W_{e^{-}\rightarrow\gamma}\!(E',E)\right.
\nonumber \\ &  &
+\left.l_{e^{+}}\!(E',X)\,
W_{e^{+}\rightarrow\gamma}\!(E',E)\right]+S_{\gamma}^{{\rm
e/m}}\!(E,X),
\end{eqnarray}
where $l_{a}\!(E,X)$ ($a=e^{-},\, e^{+},\,\gamma$) are energy spectra
of electrons, positrons, and photons at depth%
\footnote{
In the absence of particle decays there is no dependence on a particular
shower trajectory, apart from the density effect correction.} $X$, 
$\sigma_{a}\!(E)$ are interaction cross sections (in units area/mass,
see Sec.~\ref{sub:Hadronic-cascade-equations})
\begin{eqnarray}
\sigma_{e^{-}}&=&\sigma_{\rm (bremsstrahlung)}
+\sigma_{\rm (Moeller)}
\nonumber \\
\sigma_{e^{+}}&=&\sigma_{\rm (bremsstrahlung)}
+\sigma_{\rm (Bhabha)} +\sigma_{\rm (annihilation)}
\nonumber \\
\sigma_{\gamma}&=&\sigma_{\rm (pair~production)} +\sigma_{\rm (Compton)}
+\sigma_{\rm (photonuclear)} +\sigma_{\rm (muon~pair)},
\end{eqnarray}
and $W_{d\rightarrow a}(E',E)$ are corresponding differential
energy spectra of secondary particles
\begin{eqnarray}
 &  & W_{e^{-}\rightarrow e^{-}}\!(E',E)=
W_{e^{-}\rightarrow e^{-}}^{{\rm brems}}\!(E',E)+W_{e^{-}\rightarrow e^{-}}^{{\rm Moeller}}\!(E',E)
\nonumber \\
 &  & W_{e^{+}\rightarrow e^{+}}\!(E',E)=
W_{e^{-}\rightarrow e^{-}}^{{\rm brems}}\!(E',E)+W_{e^{+}\rightarrow e^{+}}^{{\rm Bhabha}}\!(E',E)
\nonumber \\
 &  & W_{\gamma\rightarrow\gamma}\!(E',E)=
W_{\gamma\rightarrow\gamma}^{{\rm Compton}}\!(E',E)
\nonumber \\
 &  & W_{e^{-}\rightarrow\gamma}\!(E',E)=
W_{e^{-}\rightarrow e^{-}}^{{\rm brems}}\!(E',E'-E)
\nonumber \\
 &  & W_{e^{+}\rightarrow\gamma}\!(E',E)=
W_{e^{-}\rightarrow e^{-}}^{{\rm brems}}\!(E',E'-E)+W_{e^{+}\rightarrow\gamma}^{{\rm annih}}\!(E',E)
\nonumber \\
 &  & W_{e^{+}\rightarrow e^{-}}\!(E',E)=
W_{e^{+}\rightarrow e^{+}}^{{\rm Bhabha}}\!(E',E'-E)
\nonumber \\
 &  & W_{\gamma\rightarrow e^{-}}\!(E',E)=
W_{\gamma\rightarrow e^{-}}^{{\rm pair}}\!(E',E)+W_{\gamma\rightarrow\gamma}^{{\rm Compton}}\!(E',E'-E)
\nonumber \\
 &  & W_{\gamma\rightarrow e^{+}}\!(E',E)=
W_{\gamma\rightarrow e^{-}}^{{\rm pair}}\!(E',E) .
\label{W-def}
\end{eqnarray}
Here $\sigma_{\rm (Moeller)}$ and $\sigma_{\rm (Bhabha)}$
correspond to the process of $\delta$-electron knock out above some
energy threshold $E_{\min}^{{\rm e/m}}$
\begin{equation}
\sigma_{e^{-/+}}^{{\rm Moeller/Bhabha}}\!(E)=
\int_{E_{\min}^{{\rm e/m}}}^{E}\! dE'\, W_{e^{-/+}\rightarrow e^{-/+}}^{{\rm Moeller/Bhabha}}\!(E,E'),
\label{moeller}
\end{equation}
whereas the contribution of those processes below $E_{\min}^{{\rm e/m}}$
is treated as continuous energy losses and constitutes a part of $\beta_{{\rm ion}}^{e^{\pm}}\!(E)$
\begin{equation}
\beta_{e^{\pm}}^{{\rm ion}}\!(E)=
\left.-\frac{dE_{e^{\pm}}}{dX}\right|_{{\rm ionization}+\delta(<E_{\min}^{{\rm e/m}})}.
\label{beta-ion}
\end{equation}

The bremsstrahlung cross section diverges due to
the characteristic infra-red singular behavior $1/E'$ of the secondary
photon spectrum 
$W_{e^{-}\rightarrow\gamma}^{{\rm brems}}\!(E,E')
=W_{e^{-}\rightarrow e^{-}}^{{\rm brems}}\!(E,E-E')$
and normaly requires to introduce some low energy cutoff. The sub-cutoff
photon emission could be treated as continuous ``radiation''
energy losses as is done in \noun{egs4} \cite{egs4}. However, 
this is not necessary in our
calculation scheme as is shown in Appendix \ref{sub:Numerical-em},
where the numerical solution is presented.


\subsection{Photonuclear effect and muon pair production}

To take into account hadron and muon production by photons, two additional
terms are introduced that couple e/m and hadronic cascade
equations via source terms. Photoproduction of hadrons, i.e.\  photonuclear
interaction, is implemented using the cross section of
Ref.~\cite{Stanev:1986cy}. Particle production distributions are
approximated by those of $\pi^0$-air interactions.
Then the corresponding source term can be written as
\begin{equation}
S_{a}^{{\rm em\rightarrow had}}(E,X)=\int_{E}^{E_{\max}}\! dE'\;
l_{\gamma}(E',X)\; W_{\pi^{0}\rightarrow
a}(E',E)\,\sigma_{\gamma}^{\mathrm{photonuc}}(E'),
\end{equation}
where $W_{\pi^{0}\rightarrow a}(E',E)$ is defined 
in analogy to Eq.~(\ref{W-def}).
With a small cross section, a photon can also produce a $\mu^{+}\mu^{-}$
pair.
This gives another contribution to the hadronic source term 
\begin{equation}
S_{\mu}^{{\rm em\rightarrow\mu}}(E,X)=\int_{E}^{E_{\max}}\! dE'\;
l_{\gamma}(E',X)\;
W_{\gamma\rightarrow\mu}(E',E)\,\sigma_{\gamma}^{\mathrm{mu-pair}}(E').
\end{equation}
Finally the source term defined in Eq.~(\ref{source1}) becomes
\begin{equation}
S_{a}^{{\rm had}}(E,X)=S_{a}^{{\rm MC\rightarrow had}}(E,X)
+S_{a}^{{\rm em\rightarrow had}}(E,X)
+S_{\mu}^{{\rm em\rightarrow\mu}}(E,X)\;\delta _a^{\mu}\,.
\end{equation}


\section{Applications\label{eas-char}}

In the following we demonstrate the reliability of the \noun{conex} code
by comparing predictions for shower observables calculated with cascade
Eqs.\ and the full hybrid scheme to that of MC simulations. As
\noun{conex} can also  run in pure MC mode, both
\noun{conex} and \noun{corsika} are used for calculating the MC
predictions.

If not specified otherwise, the
calculations were performed using the \textsc{qgsjet} model for hadronic
interactions at energies $E>E_{{\rm low}}^{{\rm had}}=80$ GeV and
\textsc{gheisha} model for $E<E_{{\rm low}}^{{\rm had}}$. Our default
choice for the cutoff between the MC and the numerical 
parts is $E_{{\rm thr}}=10^{-2}E_{0}$,
$E_{0}$ being the energy of the primary particle. The default energy
grid for solving the cascade Eqs.\ is 30 bins per energy
decade ($d_{E}=30$) for hadrons and e/m particles and  the slant depth
binning has a 5 g/cm$^{2}$ elementary step ($\Delta X$). When applying
the hybrid scheme, high energy particles are treated in MC. As a consequence
the energy transfer from hadronic to e/m particles is more precise,
allowing us to use larger bins, i.e.\ 20 bins per energy decade and
a 10 g/cm$^{2}$ slant depth step size.


\subsection{Hadronic shower component}

\begin{figure}
\begin{center}\begin{minipage}[t]{0.45\columnwidth}%
\begin{center}
\includegraphics[width=1.0\columnwidth,keepaspectratio]{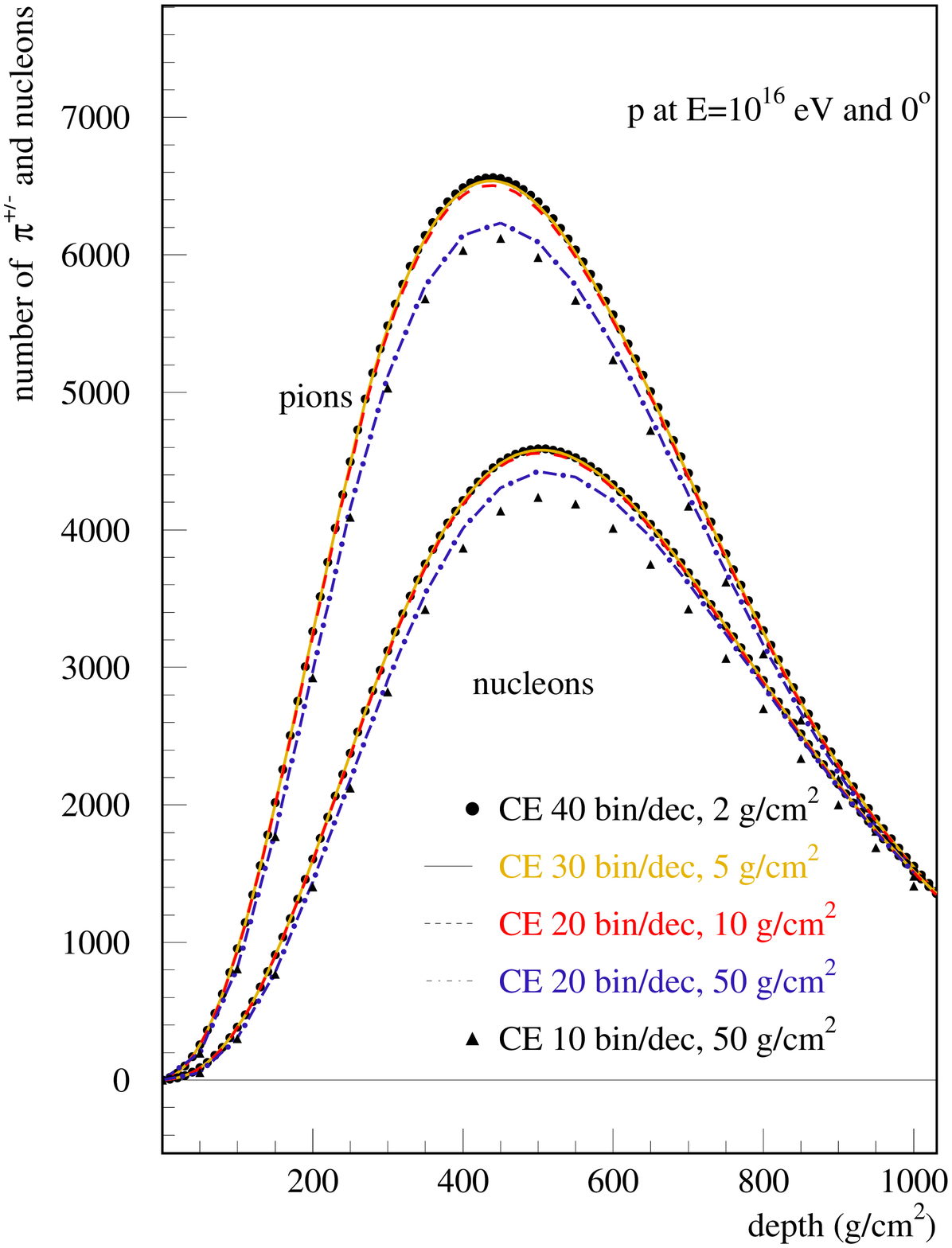}
\end{center}\end{minipage}%
\hfill
\begin{minipage}[t]{0.45\columnwidth}%
\begin{center}\includegraphics[%
  width=1.0\columnwidth,
  keepaspectratio]{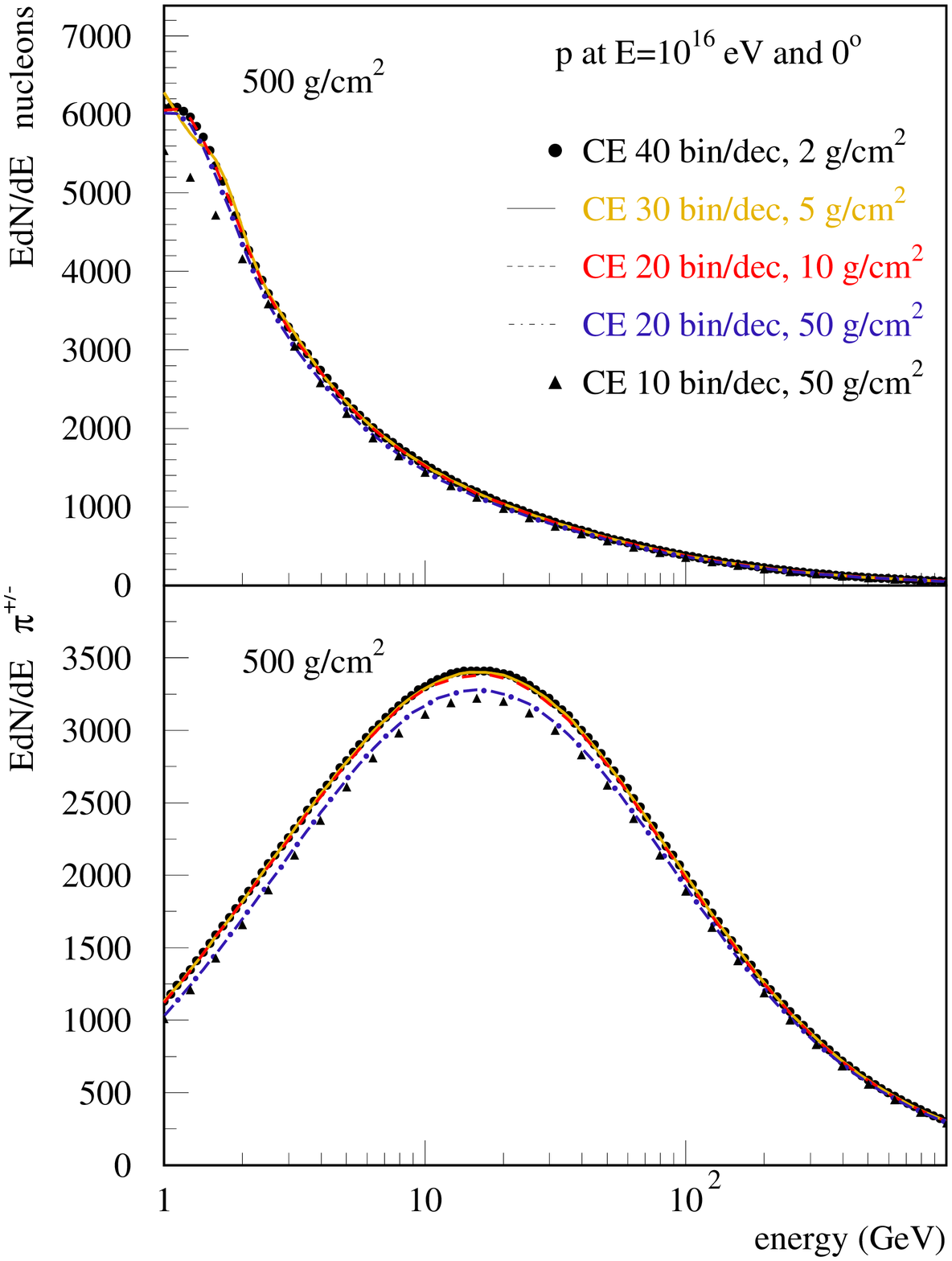}\end{center}\end{minipage}%
\end{center}
\caption{Average hadronic shower size profiles
(left panel) and energy spectra 
at $X= 500$ g/cm$^{2}$ (right panel) of nucleons and charged
pions for proton-initiated vertical ($\theta=0^{{\rm o}}$) showers
of $10^{16}$ eV. Compared are the results of 
solving numerically the system of cascade equations (CE) with different
discretization bin sizes in energy and depth.
\label{cap:Hadron-stability-16}}
\end{figure}

In Fig.~\ref{cap:Hadron-stability-16} we investigate the stability
of our scheme and compare both longitudinal profiles of nucleons and
charged pions and their energy spectra at 500 g/cm$^{2}$ for
different choices of energy and depth discretization.
Results only change significantly for very large discretization
intervals.

\begin{figure}[H]
\begin{center}\begin{minipage}[t]{0.49\columnwidth}%
\begin{center}
\includegraphics[width=1.0\columnwidth,keepaspectratio]{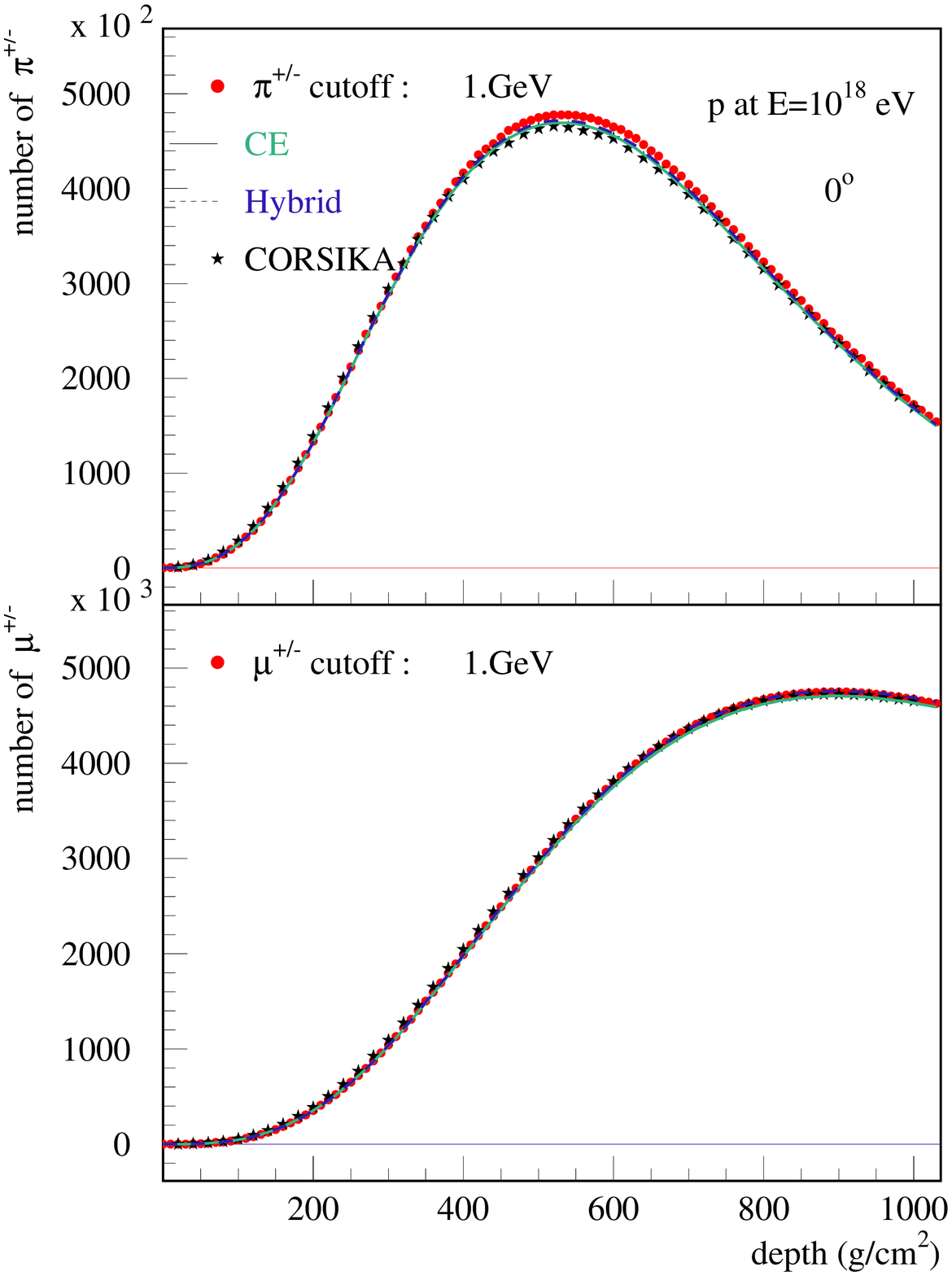}
\end{center}\end{minipage}%
\hfill
\begin{minipage}[t]{0.49\columnwidth}%
\begin{center}\includegraphics[%
  width=1.0\columnwidth,
  keepaspectratio]{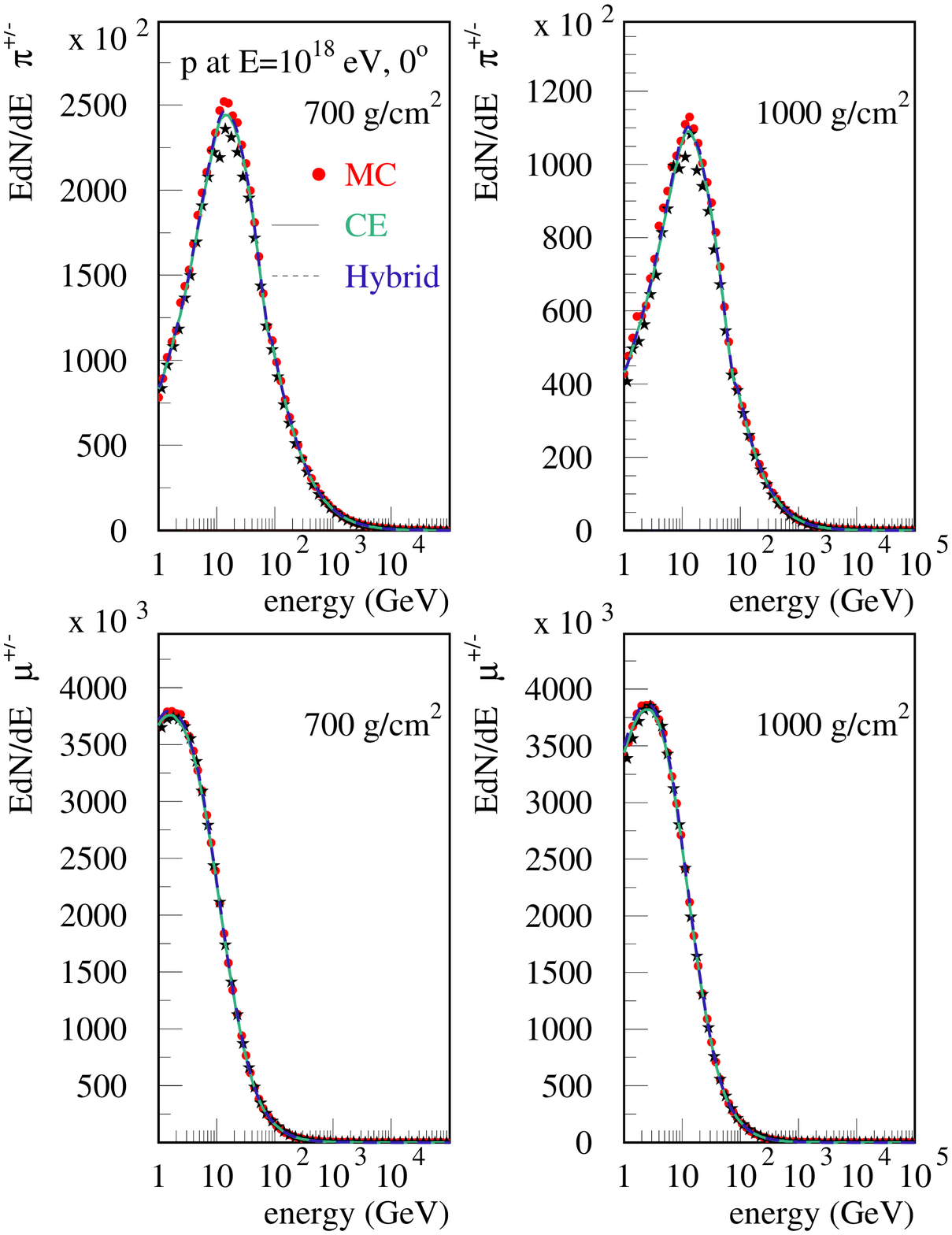}\end{center}\end{minipage}%
\end{center}

\caption{Average longitudinal shower size profiles (left panel) and energy
spectra (right panel) of charged 
pions and muons with energies
above 1 GeV. The calculations were done for proton-initiated vertical 
showers of $10^{18}$ eV. Compared are the predictions obtained with
\noun{conex} applying the hybrid
(dashed line), pure MC (points) and numerical
calculation (full line) schemes. In addition CORSIKA predictions are
shown as symbols (stars).
\label{cap:Longitudinal-profiles-of-hadron-18}}
\end{figure}

In Fig.~\ref{cap:Longitudinal-profiles-of-hadron-18}
we plot similar characteristics of charged pions and muons for $10^{18}$
eV proton-initiated showers simulated
with \noun{qgsjet 01} at high energy and \noun{gheisha}
at low energy. The results are compared to \noun{corsika} predictions.
The agreement between the results from the different 
\noun{conex} calculation methods as well as
\textsc{corsika} simulations is very good.

\subsection{Electromagnetic shower component}

\begin{figure}[H]
\begin{minipage}[t]{0.46\columnwidth}%
\begin{center}\includegraphics[%
  clip,
  width=1.0\columnwidth,
  keepaspectratio]{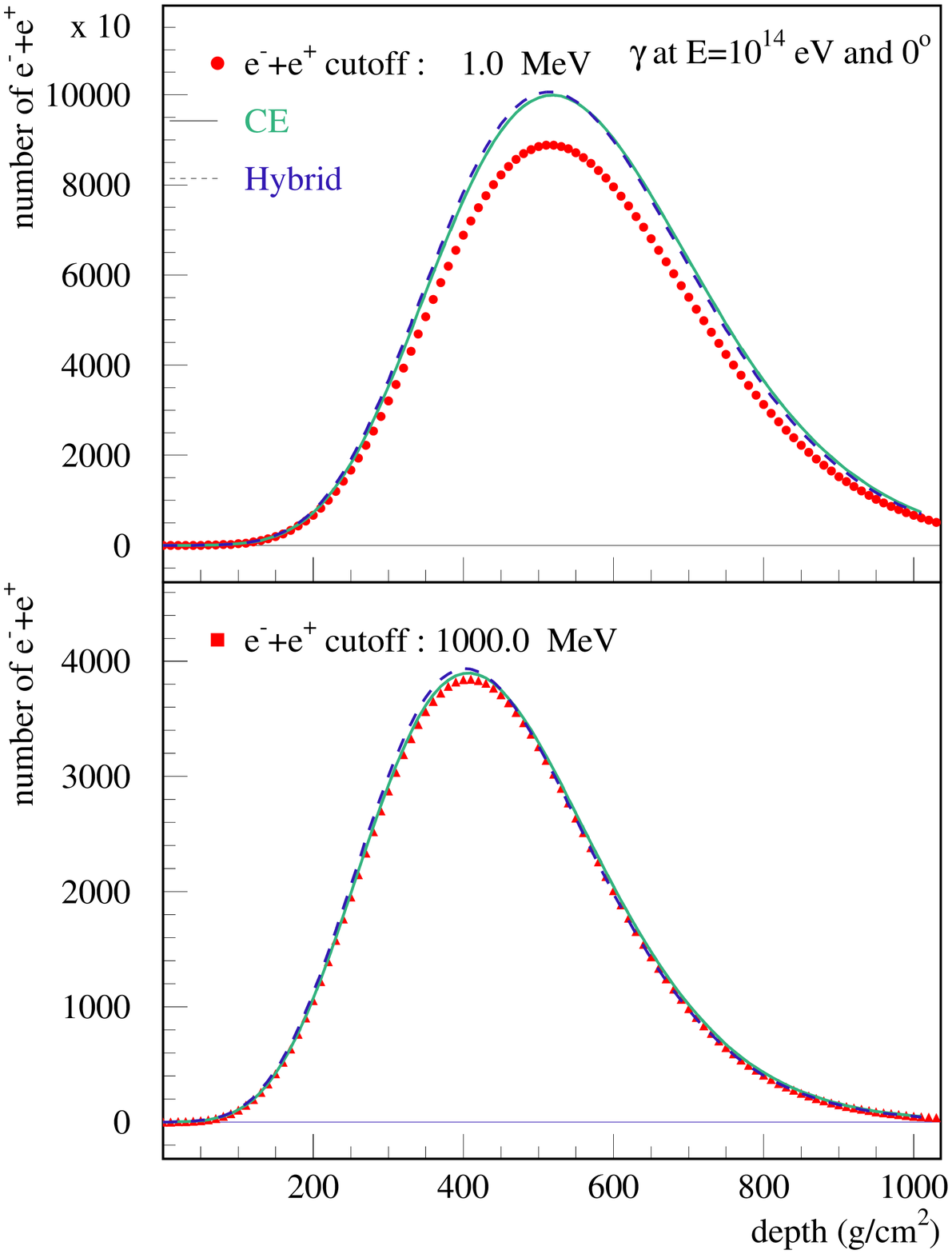}\end{center}\end{minipage}%
\begin{minipage}[t]{0.46\columnwidth}%
\begin{center}\includegraphics[%
  width=1.0\columnwidth,
  keepaspectratio]{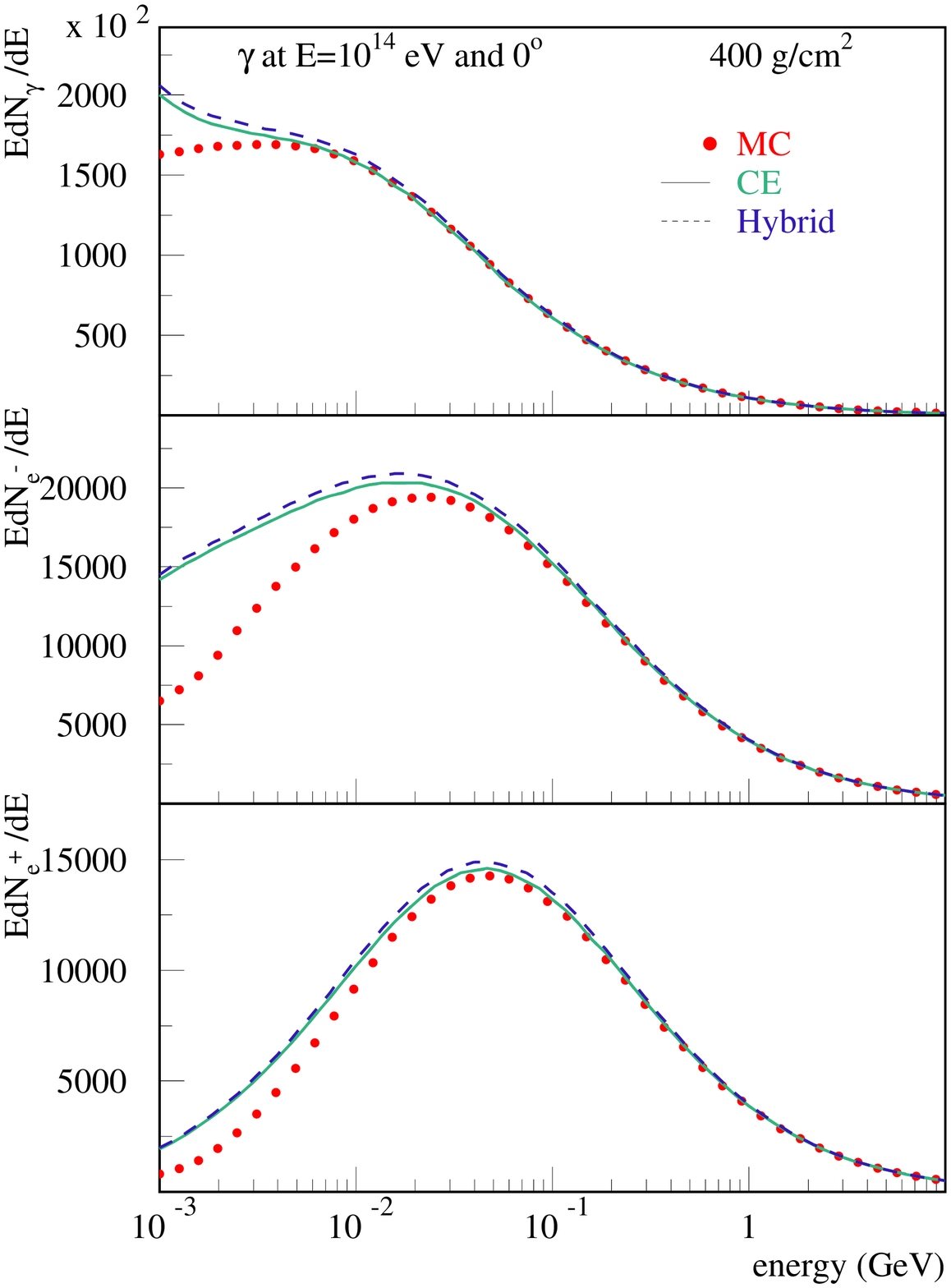}\end{center}
\end{minipage}%
\caption{
Average longitudinal shower size profiles of charged 
particles for 
$E_{\min}^{{\rm e/m}}=$1 MeV (top left panel)
and $E_{\min}^{{\rm e/m}}=$1000 MeV (bottom left panel). Particle
energy spectra of photons, electrons, and positrons are given in the
right panel. 
All curves are calculated for photon-initiated e/m showers 
using both hybrid (dashed line), pure
MC (points) and cascade equation (solid line) approaches.
\label{cap:Longitudinal-profiles-of-gamma-14-nm}
}
\end{figure}

\begin{figure}[H] 
\begin{minipage}[b]{0.46\columnwidth}%
\begin{center}
\includegraphics[clip, width=1.0\columnwidth,keepaspectratio]{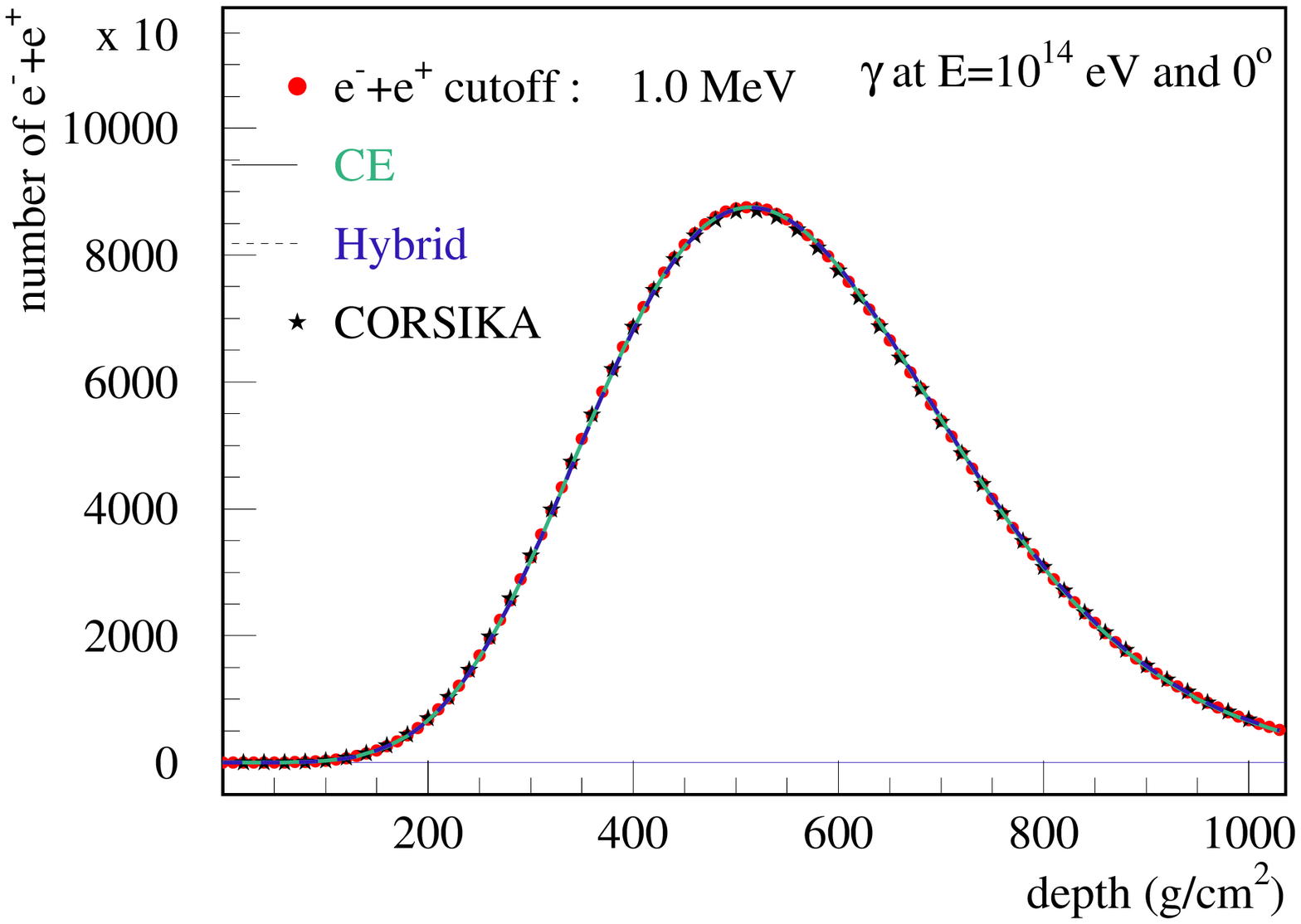}
\\
\includegraphics[clip, width=1.0\columnwidth,keepaspectratio]{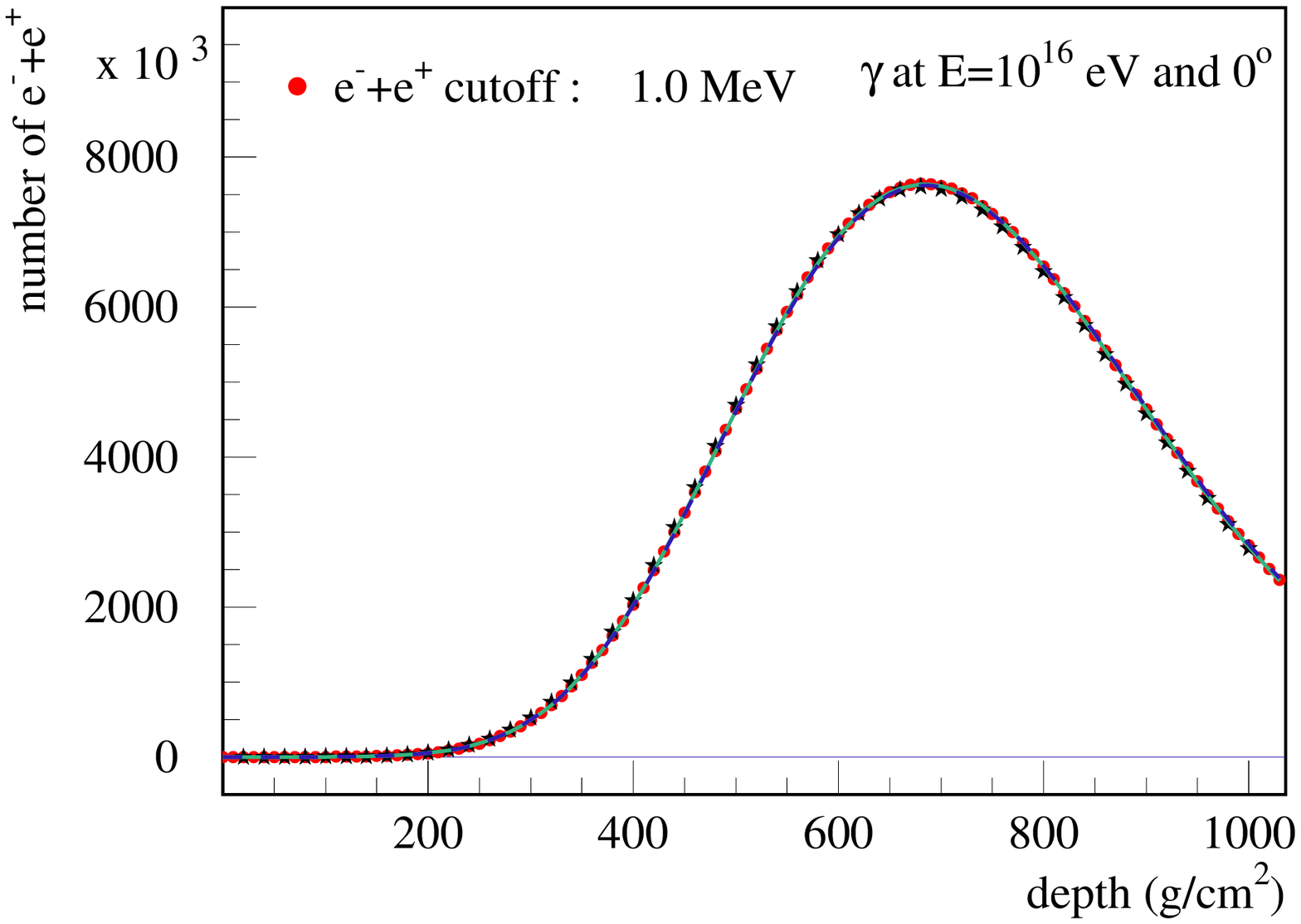}
\end{center}
\end{minipage}%
\begin{minipage}[b]{0.5\columnwidth}%
\begin{center}\includegraphics[%
  width=1.0\columnwidth,
  keepaspectratio]{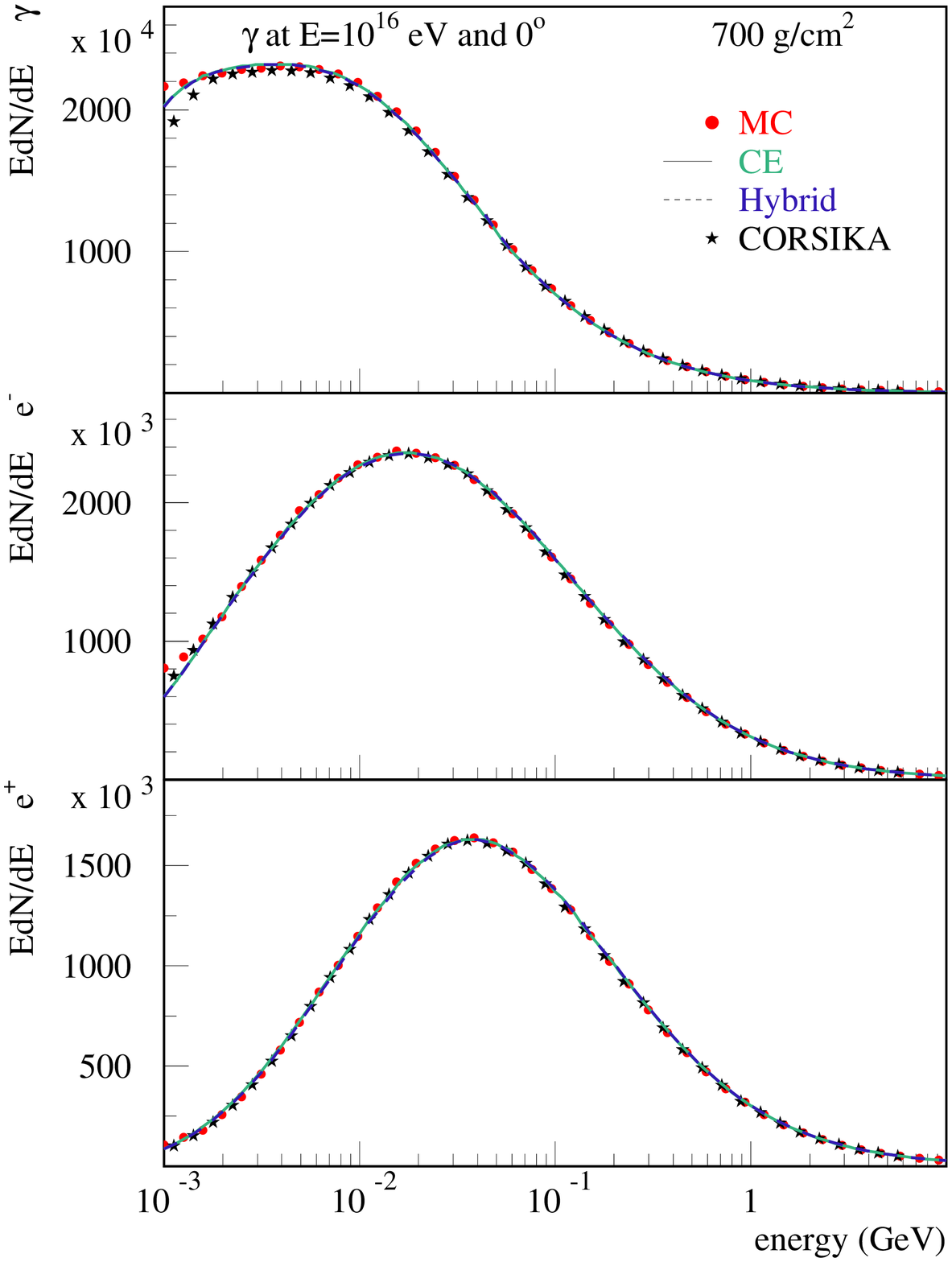}\end{center}
\end{minipage}%

\caption{Left panel: 
Average longitudinal shower size profiles of charged particles for $10^{14}$
eV (top) and $10^{16}$ eV (bottom) 
photon-initiated showers and $E_{\min}^{{\rm e/m}}=1$~MeV. 
Right panel: Particle energy spectra at $X = 700$~g/cm$^2$.
Shown are the results of MC simulations with \noun{conex} and
\noun{corsika} and of the hybrid and cascade equation schemes.
Here an ``average angular deflection'' is used in
the numerical and hybrid schemes.
\label{cap:Longitudinal-profiles-of-gamma-14}}
\end{figure}

The longitudinal profiles of electrons, positrons, and photons for
a $10^{14}$ eV vertical photon-initiated
shower are shown in Fig.~\ref{cap:Longitudinal-profiles-of-gamma-14-nm}
(left panel).
The shower size profiles are given for the cutoff energies
$E_{\min}^{{\rm e/m}}=$1 MeV,
1000 MeV using again the hybrid, pure MC, and cascade equation approaches.
While for a large cutoff energy, for example 1000 MeV, 
the agreement between the different methods
is good we notice systematically larger particle numbers in the hybrid
and numerical calculations for $E_{\min}^{{\rm e/m}}=$1 MeV.
The corresponding difference is clearly visible 
in the particle energy spectra and is related to spatial effects
in the shower development. Low energy electrons (positrons) undergo
significant angular deflections due mainly to multiple Coulomb scattering.
In turn low energy bremsstrahlung photons produced by such deflected
particles also have significant directional deviations from the initial
shower axis. If only the track projected to the shower axis is
considered, this leads to an apparently faster absorption of low energy
particles (higher interaction rate and ionization energy loss)
compared to that expected for particles traveling along the
shower axis only (see also discussion in \cite{Alvarez-Muniz:2003my}).
Although a full account of this effect requires a three-dimensional
treatment of the particle cascade at MeV energies a reasonable improvement
can be achieved by introducing an ``average angular deflection''. As
the effect is only important for low energy leptons which anyway lose
their energy quite fast we may estimate the corresponding average
scattering angle of an electron (positron) as
\begin{equation}
\langle\theta^{2}\rangle\sim\frac{E_{s}^{2}}{E^{2}}L\!(E),
\label{theta-scat0}
\end{equation}
where $E_{s}\simeq21$ MeV and $L\!(E)$ is the average travel distance
of an electron of energy $E$ in units of radiation length. 
With $L\!(E)\simeq E/E_{{\rm crit}}$ and
$E_{{\rm crit}}\simeq81$\,MeV, we have
\begin{equation}
\langle\theta^{2}\rangle\sim\frac{E_{s}^{2}}{ E\, E_{{\rm crit}}}\,.
\label{theta-scat}
\end{equation}
For numerical applications, expression (\ref{theta-scat}) has to be
modified to satisfy the boundary condition $\theta \le \pi/2$. In the
following we chose the ansatz
\begin{equation}
\langle\theta^{2}\rangle=
\left(\frac{\pi}{2}\right)^{2}\left[1-\exp\!\left(-\frac{E_{e^{\pm}}^{{\rm eff}}}{E}\right)\right],
\label{theta-scat-final}
\end{equation}
which reduces to the functional form of (\ref{theta-scat}) for 
large $E$ and approaches $(\pi/2)^{2}$ in the small $E$ limit.
The same kind of formula is used for photons
but with a different parameter $E_{\gamma}^{{\rm eff}}$, as photons
themselves do not undergo multiple scattering. Good
agreement between fully three-dimensional MC simulations and cascade
equation calculations is obtained for 
$E_{e^{\pm}}^{{\rm eff}}=9.5\cdot10^{-4}$~GeV
and $E_{\gamma}^{{\rm eff}}=5\cdot10^{-4}$~GeV. This is shown
in Fig.~\ref{cap:Longitudinal-profiles-of-gamma-14} where the shower
size profiles and energy spectra of the two approaches are compared
for e/m showers of $10^{14}$ eV and $10^{16}$ eV.

\begin{figure}[H] 
\begin{minipage}[b]{0.46\columnwidth}%
\begin{center}
\includegraphics[clip, width=1.0\columnwidth,keepaspectratio]{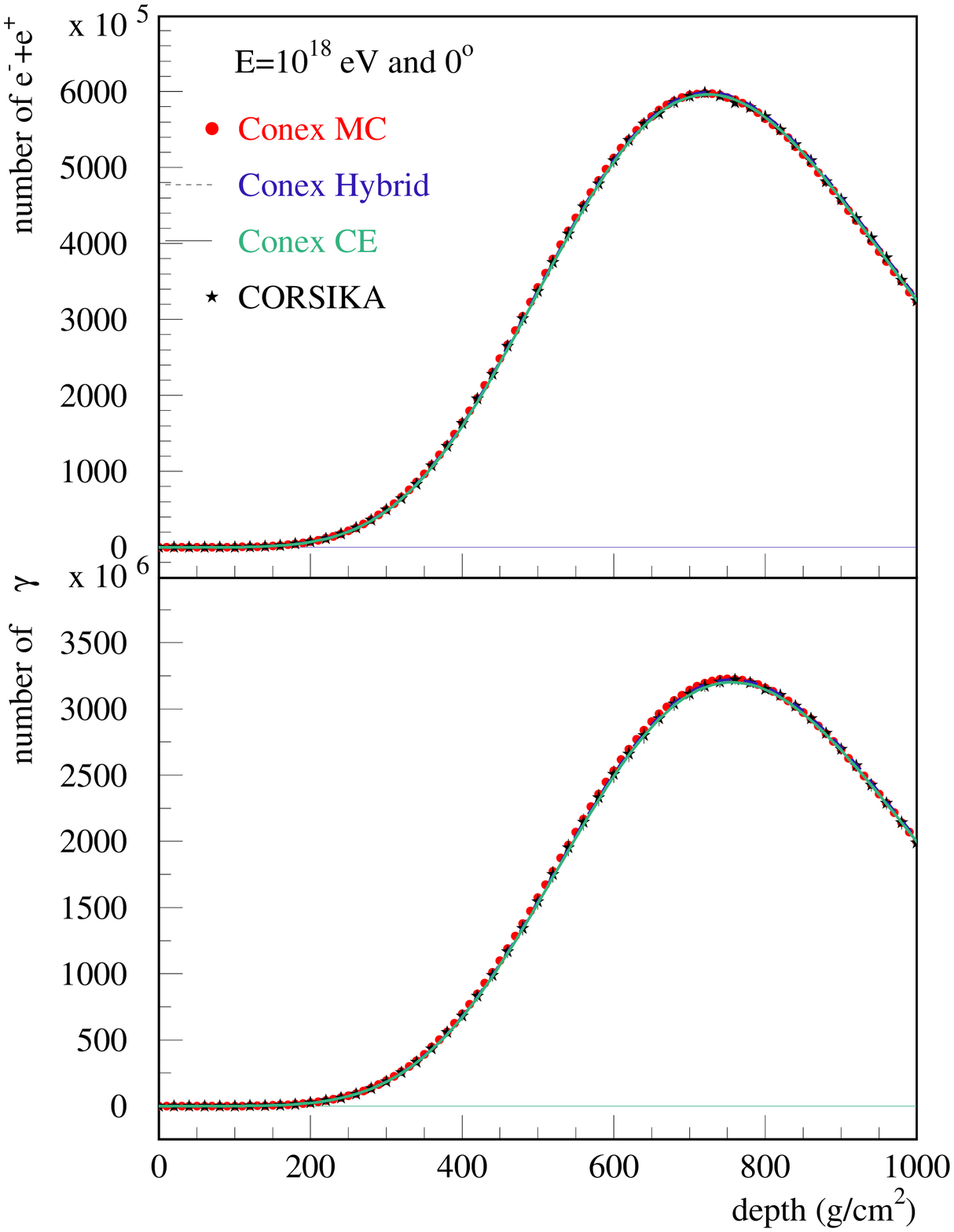}
\end{center}
\end{minipage}%
\hfill
\begin{minipage}[b]{0.5\columnwidth}%
\begin{center}\includegraphics[width=0.94\columnwidth,keepaspectratio]{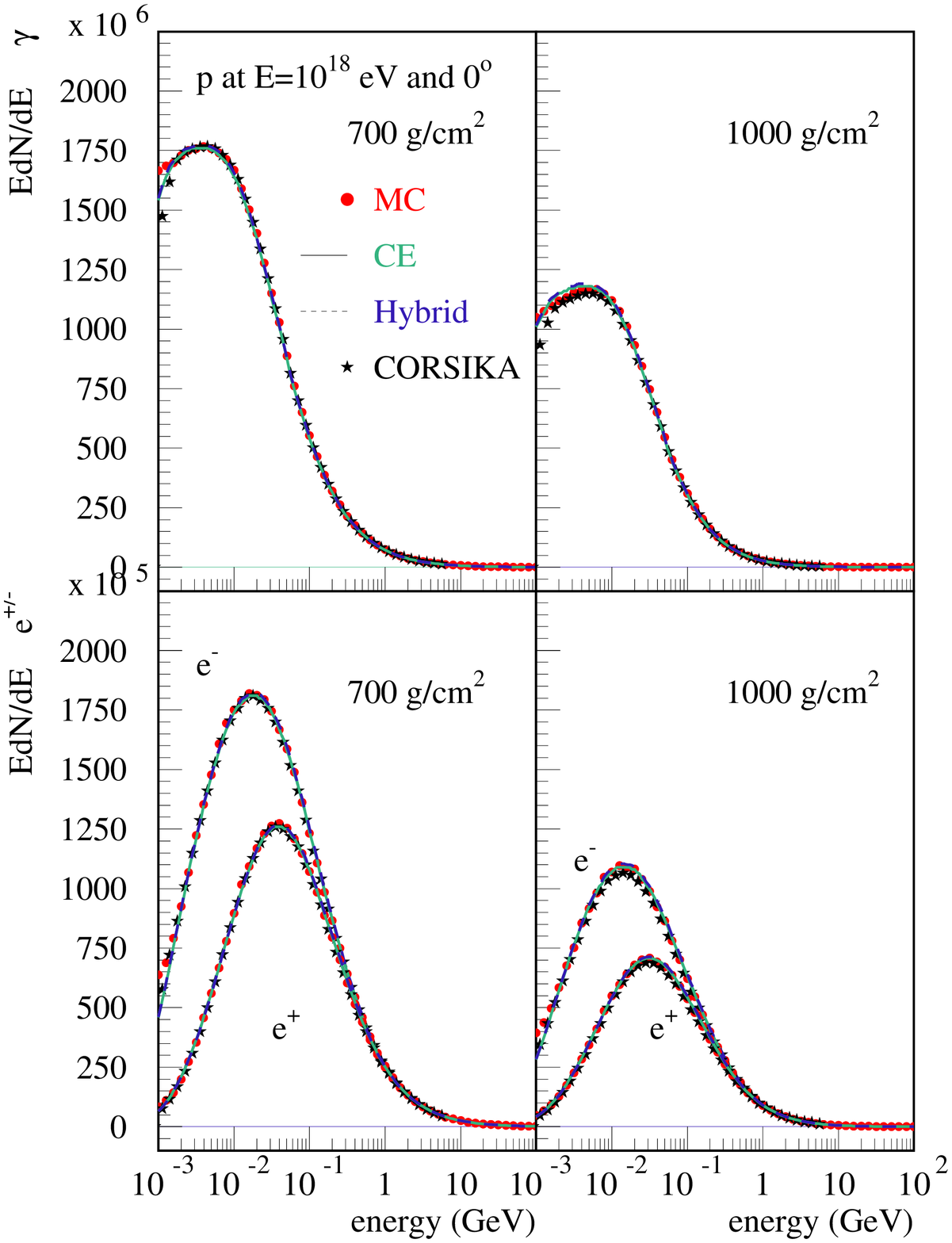}
\end{center}
\end{minipage}%

\caption{
Left panel: Average longitudinal profiles of charged particles and photons of
energies above 1 MeV for proton-initiated vertical
showers of $E_{0}=10^{18}$ eV. 
 Right panel: Particle energy spectra of photons, electrons, and positrons for
the atm.\ depths $X= 700$ and 1000 g/cm$^{2}$. Shown are the results from the hybrid
calculation (dashed line), pure MC simulation (points), and numerical cascade Eqs.\
 solution
(full line). In addition \textsc{corsika} predictions are given by stars.
\label{cap:p-charged-18}
}
\end{figure}

After having shown that both hadronic and e/m showers can be well described by the presented
cascade equations and the corresponding hybrid simulation scheme, we test the coupling between 
the e/m and hadronic shower components.   
In Fig.~\ref{cap:p-charged-18}
we compare both longitudinal profiles and energy spectra of e/m particles
for vertical proton-induced showers of $10^{18}$ eV. Very good agreement is found between the results
of the different calculation methods and also with the \textsc{corsika} predictions. The simulations 
were performed using \textsc{qgsjet} 01 as high-energy interaction model and \textsc{gheisha} 
for hadronic interactions at low energy.


\subsection{Energy deposit}

The longitudinal profile of simulated EAS depends on the e/m low-energy cut-off $E_{\rm min}^{\rm e/m}$ 
used in the calculation (in this work 1 MeV). Lowering this threshold to, for example, 50 keV
would increase the number of charged particles at the shower maximum by 
a few percent \cite{Nerling:2006yt}. In the case of the photon number, the dependence of the 
simulation results on the low-energy cut-off is much stronger. In fact, the number of photons diverges 
if the simulation cut-off is set to 0. Problems of this kind can be avoided if, instead of the 
secondary particle profile, the ionization energy deposit profile of showers is considered. 
Another advantage of the energy deposit profile is its direct relation to the light curve 
measured in air fluorescence experiments. Current measurements support the theoretical expectation that
the fluorescence yield is proportional to the ionization energy deposit \cite{Arciprete:2006pb}.

Therefore, the hybrid simulation scheme implemented in \textsc{conex}
has been extended to also allow the calculation of energy deposit profiles as described below.  

The calculation of the energy deposit profile in the MC part of \textsc{conex} is very similar to that 
in \textsc{corsika}, see \cite{Risse:2003fw}. The
deposited ionization energy is counted for each particle and traversed slant
depth bin. If a particle reaches the
cut-off energy of the calculation, depending on the particle type, 
either all its energy or a part of it is assumed to be deposited locally.
 For example, a positron is expected to annihilate in the 
end and, therefore, all  its  energy plus electron  mass are assumed 
to be deposited. The different energy contributions taken to be deposited locally are given in 
Table~\ref{cap:Deposited-energy}.

\begin{table}

\caption{
Energy deposited locally for particles with $E_{\rm kin}<E_{\rm min}$ ($\mathrm{m_{e}}$ denotes
the electron mass and $m_{N}$\ is the nucleon mass).
\label{cap:Deposited-energy}
}
\begin{center}
\begin{tabular}{|c|c|c|c|c|c|c|c|c|}
\hline 
particle &
$\gamma$&
$e^{-}$&
$e^{+}$&
$\mu^{\pm}$&
$\nu$&
protons&
neutrons&
$\pi^{\pm}$,$\mathrm{K^{\pm}}$\tabularnewline
\hline 
energy deposit&
$E_{\rm tot}$&
$E_{\rm kin}$&
$E_{\rm tot}+m_{e}$&
$E_{\rm tot}/3$&
0&
$E_{\rm kin}$&
0&
$E_{\rm tot}/4$\tabularnewline
\hline
\end{tabular}

\begin{tabular}{|c|c|c|c|c|c|}
\hline 
particle &
$\mathrm{K^{0}}$&
baryons&
anti-baryons&
nuclei&
anti-nuclei\tabularnewline
\hline 
energy deposit&
$E_{\rm tot}/2$&
$E_{\rm tot}-m_{N}$&
$E_{\rm tot}+m_{N}$&
$E_{\rm tot}-A\cdot m_{N}$&
$E_{\rm tot}+A\cdot m_{N}$\tabularnewline
\hline
\end{tabular}
\end{center}
\end{table}

Solving cascade equations numerically one can calculate the energy deposit of particles with
$E>E_{\rm min}$ explicitly. In addition the number of newly created particles at each step 
in atmospheric depth is given by the source functions. However,
the energy carried by particles falling below the cut-off energy threshold and that 
of neutrinos is not directly calculated. 

For the e/m cascade equations, the total energy 
deposit per slant depth bin can be estimated from energy conservation. The deposited 
energy follows from the 
difference between the total kinetic energy at slant depths $X_m$ and $X_{m-1}$, including a 
correction for positrons similar to the MC treatment
\begin{eqnarray}
E_{\rm dep}^{\rm e/m}(X_{m}) & = & \sum_{i=i_{\rm min}}^{i_{\rm max}}
\left(E_{i}^{\mathrm{e/m}}\sum_{a=e^{\pm},\gamma}l_{a}^{i}(X_{m})
+2\, m_{\mathrm{e}}\;l_{e^{+}}^{i}(X_{m})\right)
\nonumber\\
 & &-  \sum_{i=i_{min}}^{i_{max}}
\left(E_{i}^{\mathrm{e/m}} \sum_{a=e^{\pm},\gamma} l_{a}^{i}(X_{m-1})
+2\, m_{\mathrm{e}}\;l_{e^{+}}^{i}(X_{m-1})\right),
\end{eqnarray}
where $m_{e}$\ is electron mass.

A similar energy conservation-based method can be applied to the hadronic
shower component, however, the deposited
energy and the energy going into neutrino production have to be distinguished.
Since we know the ionization energy deposited by hadrons and can calculate
the number of neutrinos produced within a given slant depth bin, we
can obtain the energy of all the particles falling below the low-energy threshold from
\begin{equation}
E_{\rm cut}(X_{m})=E_{\rm bal}(X_{m})-E_{\rm ion}(X_{m})-E_{\nu}(X_{m}),
\end{equation}
where
\begin{eqnarray}
E_{\rm bal}(X_{m}) & = & \sum_{i=i_{min}}^{i_{max}}\left(E_{i}^{\mathrm{had}}
\sum _a h_{a}^{i}(X_{m})+\sum_{a}m_{a}\; h_{a}^{i}(X_{m})\right)\nonumber\\
 & - & \sum_{i=i_{min}}^{i_{max}}\left(E_{i}^{\mathrm{had}}
\sum _a h_{a}^{i}(X_{m-1})+\sum_{a}m_{a}\; h_{a}^{i}(X_{m-1})\right)\\
E_{\rm ion}(X_{m}) & = & \sum_{i=i_{min}}^{i_{max}}\sum_{a}\, h_{a}^{i}(X_{m-1})\:
\left(1-\exp\left[-\frac{\beta_{a}^{{\rm ion}}(E_{i}^{{\rm had}})}{
E_{i}^{{\rm had}}-E_{i-1}^{{\rm had}}}\Delta X\right]\right)\nonumber\\
 & \times & \exp\!\left[-\frac{1-W_{a\rightarrow a}^{ii}}{\lambda_{a}(E_{i}^{{\rm had}})}\Delta X
-\frac{m_{a}\left|L(X_{m})-L(X_{m-1})\right|}{c\tau_{a}^{(0)}E_{i}^{{\rm had}}}\right]\\
E_{\nu}(X_{m}) & = & \int_{X_{m-1}}^{X_{m}}\! d X^\prime \sum_{d}\sum_{j=i+1}^{i_{\max}}\, 
h_{d}^{j}(X')\, D_{d\rightarrow\nu}^{ji}\frac{m_{d}/(c\tau_{d}^{(0)})}{\rho_{{\rm air}}(X')}.
\end{eqnarray}
To account for the fact that only a part of the energy carried by hadrons 
falling below the low-energy threshold is deposited as ionization energy,
a factor $f_{\mathrm{had}/\mu}=0.45$ is introduced \cite{Barbosa:2003dc}. 
Thus, the energy deposit for the
hadronic part of the system of cascade Eqs.\ is given by 
\begin{equation}
E_{\rm dep}^{\rm had}(X_{m})=E_{\rm ion}(X_{m})+f_{\mathrm{had}/\mu}\; E_{\rm cut}(X_{m}).
\end{equation}

\begin{figure}[H]
\begin{center}
\includegraphics[clip,width=0.55\columnwidth]{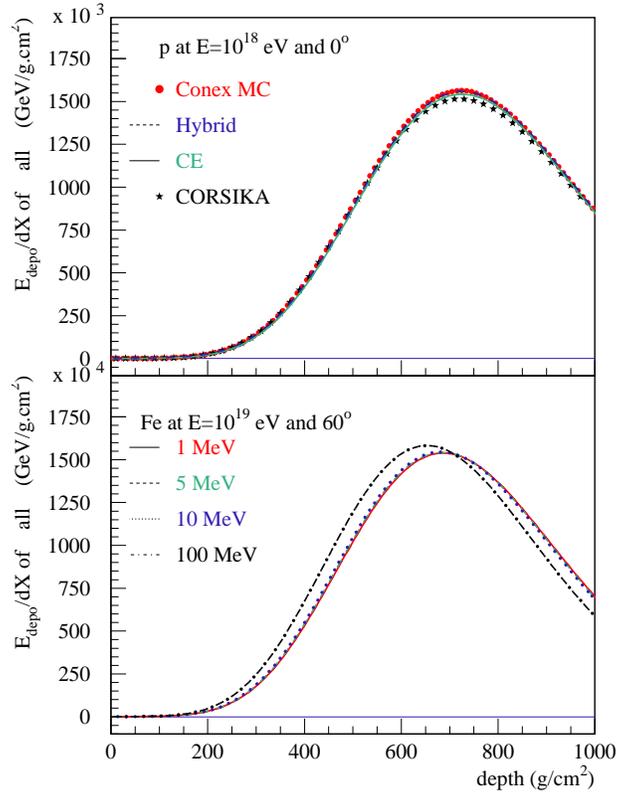}
\end{center}
\caption{
Upper panel: Average longitudinal energy deposit profile for vertical
proton-initiated showers of  $10^{18}$\,eV. The results obtained with the
e/m energy cutoff $E_{\rm min}^{{\rm e/m}}=1$  MeV, using hybrid
(dashed line), MC (points), or numerical approaches (full
line), are  compared to \noun{corsika} predictions (stars).
Lower panel:  Energy deposit profiles for $10^{19}$
eV  inclined ($\theta=60^{{\rm o}}$) iron-initiated showers for the
 e/m energy cutoff $E_{\min}^{{\rm e/m}}=$1 (full line),
5 (dashed line), 10 (dotted line) and 100 MeV (dashed-dotted line).
\label{Longitudinal-Energy-Deposit}}
\end{figure}

Summing the different contributions, the longitudinal
energy deposit profile can be calculated. A comparison of the results of 
different calculation methods within \noun{conex} with \noun{corsika} predictions
is shown in Fig.~\ref{Longitudinal-Energy-Deposit} (top panel). Good agreement is found. 

An interesting consistency check of the energy deposit calculation scheme 
 is the investigation of the dependence of corresponding results
  on the low-energy threshold used in the e/m cascade Eqs.  
In Fig.~\ref{Longitudinal-Energy-Deposit} (bottom panel), the mean longitudinal 
energy deposit profiles of iron-initiated showers of $\theta = 60^\circ$ and  $E = 10^{19}$\,eV are 
shown for different low-energy thresholds. The profiles are independent of this 
threshold over a wide energy range -- 
the approximation of local energy deposit breaks down 
only for a threshold energy of 100 MeV and higher.

\subsection{Shower fluctuations}

So far we have only studied mean shower observables, i.e.\ ones averaged over 
many showers. As a proper description of shower fluctuations is of central 
importance for the analysis of experimental data, we will also discuss the 
treatment of shower fluctuations.

Running \noun{conex} in hybrid mode allows us to benefit from the fast 
numerical solution of cascade Eqs.\ and, at the same time, to obtain a good
description of shower-to-shower fluctuations. Here,
the key parameter of the method is the energy threshold that separates
the explicit MC simulation from the application of cascade Eqs. By default, this energy 
threshold is set to $E_{\rm thr} = 0.01\cdot E_0$ for all particles in \noun{conex}. 
In principle, it can be chosen  differently for e/m and hadronic particles 
to further reduce the simulation time needed for high-energy showers.

\begin{figure}[tbph]
\begin{center}
\includegraphics[%
  width=0.70\columnwidth,
  keepaspectratio]{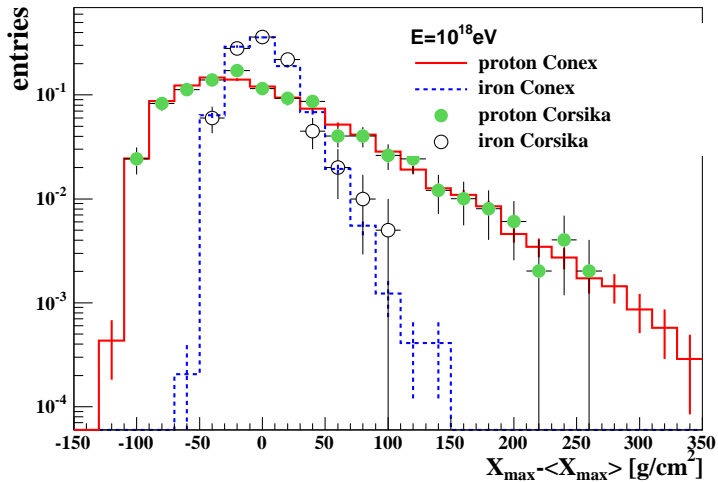}
\end{center}
\caption{Fluctuations of the shower maximum depth $\mathrm{X}_{\mathrm{max}}$around
the mean shower maximum depth $\left\langle \mathrm{X}_{\mathrm{max}}\right\rangle $
for a primary energy of $10^{18}$eV for proton and iron-initiated
showers simulated with \noun{conex} (lines) and compared with \noun{corsika}
results (points).
\label{Xmax18-fluctuations}
}
\end{figure}
In Fig.~\ref{Xmax18-fluctuations} the distribution of the depth of shower maximum 
is shown for proton- and iron-initiated showers. Within the statistical uncertainties,
the \noun{conex} results agree very well 
with that obtained with \noun{corsika} MC simulations. Not only the fluctuations but 
also the mean depth of shower maximum obtained with the two codes are the same (not shown).

\begin{figure}[tbph]
\begin{center}
\includegraphics[%
  width=0.70\columnwidth,
 keepaspectratio]{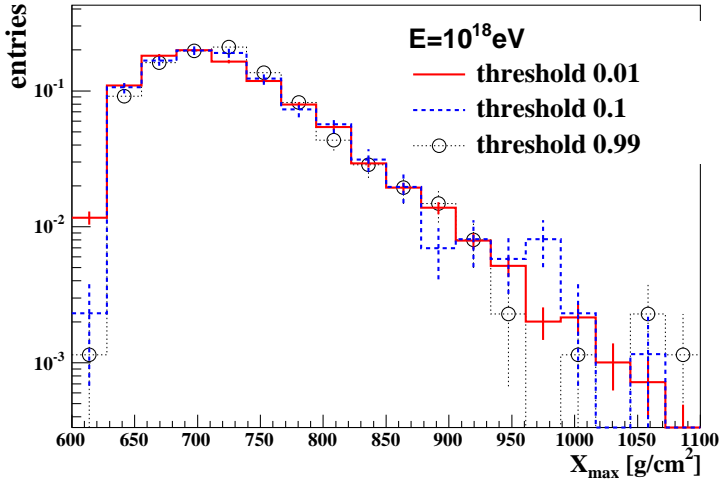}
\end{center}
\caption{Fluctuations of the shower maximum depth $\mathrm{X}_{\mathrm{max}}$
for  $10^{18}$eV  proton-initiated showers
simulated with \noun{conex} for three different energy thresholds 
$E_{\rm thr}/E_0=0.01$, 0.1, 0.99.
\label{Xmax18-threshold}
}
\end{figure}

The dependence of the simulated fluctuations on the energy threshold is shown 
in Fig.~\ref{Xmax18-threshold}. Even a threshold as large 
as $E_{\rm thr} = 0.99\cdot E_0$ is sufficient to reproduce almost the full $X_{\rm max}$ 
distribution. This comparison demonstrates that the fluctuations of
 particle production in 
the first interaction of such high-energy showers determine
 almost the entire shower profile.



\section{Summary\label{sec:Summary}}

We have developed a fast and efficient one-dimensional hybrid simulation scheme for ultra-high 
energy air showers. It combines explicit MC simulation of high-energy particle interaction, 
propagation and decay with the numerical solution of a system of cascade equations 
for calculating the low-energy part of the particle cascade. 

The presented hybrid simulation scheme is implemented in the code 
\noun{conex}\footnote{The \noun{conex} code is available upon request from Tanguy.Pierog@ik.fzk.de}. 
Several high- and low-energy hadronic interaction models are available within \noun{conex} 
to study theoretical predictions and the model-dependence of data analyses. 

All relevant interaction 
and decay processes are considered in both the MC and the cascade equation parts of \noun{conex}. 
These processes also include muon pair production and photonuclear interactions of muons. At 
ultra-high energy, the LPM effect and possible e/m preshowering in the geo-magnetic 
field are simulated.

The hybrid simulation scheme has been extended to include the simultaneous calculation of
both shower size profiles of various particles and the generation of ionization
energy deposit profiles. The latter are independent of the low-energy cut-off that has to be applied in 
all shower simulations. Knowing both the shower size profile (with an arbitrary low-energy 
cut-off) and the energy deposit profile allows us to simulate directly the fluorescence and 
Cherenkov light signal of air showers. Together with the fully three-dimensional implementation of 
the shower axis geometry, this makes \noun{conex} ideally suited for event 
simulation and data analysis of fluorescence light experiments such as 
HiRes \cite{Abbasi:2004nz}, Auger \cite{Abraham:2004dt}, TA \cite{Kasahara:2005pk}, 
and EUSO \cite{Pallavicini:2003we}.

In developing \noun{conex}, particular emphasis is put on the accuracy and reliability 
of the shower simulation to make the code directly applicable to data analysis of air shower experiments.
Extensive comparisons with \noun{corsika} simulations show that all shower distributions agree 
very well. Both mean shower profiles and energy distributions as well as their 
fluctuations were compared, only a small fraction of which could be shown in this paper.

In a forthcoming work we will study the influence of different hadronic interaction models on air shower
predictions. In particular we will investigate the total calorimetric energy deposited by a shower
in air. First results of this work have been presented in \cite{Pierog:2005aa}.

\section{Appendix}

\subsection{Geometry}

\begin{figure}[tbph]
\begin{center}
\includegraphics[%
  width=0.50\columnwidth]{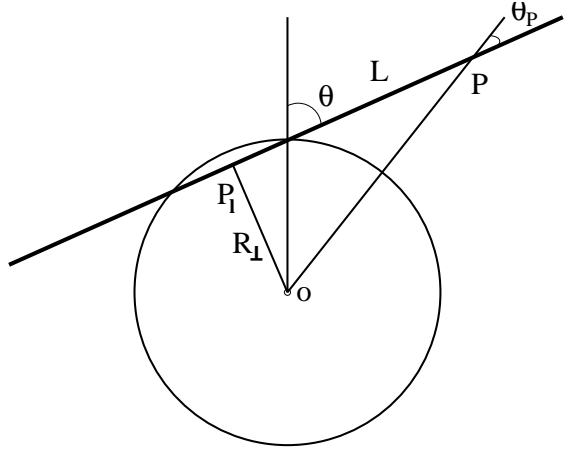}
\end{center}
\caption{Geometry for defining the shower trajectory accounting 
for the curvature of the Earth.
\label{traj}
}
\end{figure}

Using a one-dimensional treatment of EAS development, a given shower
trajectory may be characterized by a single parameter - its distance
to the center of the Earth $O$: $R_{\perp}^{T}=\left|P_{l}^{T}O\right|$
- see Fig.~\ref{traj}, where $P_{l}^{T}$ is the lowest trajectory
point. %
In case the observer is positioned at some height $h_{{\rm obs}}$
above  sea level and $R_{{\rm Earth}}+h_{{\rm obs}}>R_{\perp}^{T}$
($R_{{\rm Earth}}$ being the Earth radius) the observed shower
inclination is
\begin{equation}
\theta=\arcsin\frac{R_{\perp}^{T}}{R_{{\rm Earth}}+h_{{\rm obs}}}.
\label{theta}
\end{equation}

The position of a particle moving along a given trajectory may be
then characterized by its local azimuthal angle $\theta_{P}^{T}$
with respect to the vertical direction, or alternatively, by its height
$H_{P}^{T}$, or by the distance $L_{P}^{T}=\left|P_{l}^{T}P\right|$
to the lowest trajectory point
\begin{eqnarray}
 &  & H_{P}^{T}=\frac{R_{\perp}^{T}}{\sin\theta_{P}^{T}}
\label{height}\\
 &  & L_{P}^{T}=\frac{R_{\perp}^{T}}{\cos\theta_{P}^{T}}.
\label{dist}
\end{eqnarray}

For the solution of cascade equations, it is more convenient to use
instead the slant grammage $X_{P}^{T}$, i.e.\ the integral over the
atmospheric density $\rho_{{\rm air}}(h)$ along the trajectory $T$
from the point $P$ to infinity
\begin{equation}
X_{P}^{T}=\int_{L_{P}^{T}}^{\infty}\! dl'\;\rho_{{\rm air}}\left(H(l')\right),
\label{depth}
\end{equation}
and characterize an arbitrary particle position by two variables, 
$R_{\perp}^{T}$ and $X$.

\subsection{Numerical treatment of hadronic cascade equations\label{sub:Numerical-had}}

In order to reduce the problem to the solution of a system of ordinary
differential equations, we perform a discretization of particle energy
spectra. Introducing an energy binning as 
$E_{i}^{{\rm had}}=E_{\min}^{{\rm had}}\; C^{i-1}$,
$C=\,10^{1/d_{E}}$ ($E_{\min}^{{\rm had}}=1$ GeV, $d_{E}=10\div30$)
and replacing the smooth particle spectra $h_{a}\!(E,X)$ by discrete
contributions $h_{a}^{i}\!(X)$ of representative particles of energies
$E_{i}^{{\rm had}}$ ($i=1,...,i_{\max}$)
we may get instead of (\ref{sys1-had})
\begin{eqnarray}
\frac{dh_{a}^{i}(X)}{dX}&=&
-h_{a}^{i}(X)\left[\frac{1}{\lambda_{a}(E_{i}^{{\rm had}})}
+\left|\frac{dL}{dX}\right|\frac{m_{a}}{c\,\tau_{a}^{(0)}\:E_{i}^{{\rm had}}}+
\frac{\beta_{a}^{{\rm ion}}(E_{i}^{{\rm had}})}{E_{i}^{\mathrm{had}}-E_{i-1}^{\mathrm{had}}}\right]
\nonumber\\
 &  & +\sum_{d}\sum_{j=i}^{i_{\max}}\; h_{d}^{j}(X)\;
 \left[\frac{W_{d\rightarrow a}^{ji}}{\lambda_{d}(E_{j}^{{\rm had}})}+
 D_{d\rightarrow a}^{ji}\frac{m_{d}/(c\,\tau_{d}^{(0)})}
 {E_{j}^{{\rm had}}\;\rho_{{\rm air}}(X)}\right]
\nonumber \\
 &  & +h_{a}^{i+1}(X)\,\frac{\beta_{a}^{{\rm ion}}(E_{i+1}^{{\rm had}})}
 {E_{i+1}^{\mathrm{had}}-E_{i}^{\mathrm{had}}}+S_{ai}^{{\rm had}}(X),
\label{sys2-had}
\end{eqnarray}
where we used (\ref{dldx}), replaced the integral over parent particle
energies $\int\! dE'$ by the discrete sum $\sum_{j}$, introduced
the discretized source term $S_{ai}^{{\rm had}}(X)$ and the discrete
interaction (decay) spectra $W_{d\rightarrow a}^{ij}$ ($D_{d\rightarrow a}^{ij}$)
via
\begin{eqnarray}
S_{ai}^{{\rm had}}(X)&=&\int_{\max\left[E_{i-1}^{{\rm had}},E_{\min}^{{\rm had}}
\right]}^{E_{i}^{{\rm had}}}\! dE'\;
 S_{a}^{{\rm had}}(E',X)\; K(E'/E_{i}^{{\rm had}})
\nonumber\\
 &  & +\int_{E_{i}^{{\rm had}}}^{\min\left[E_{i+1}^{{\rm had}},
 E_{{\rm thr}}\right]}\! dE'\; S_{a}^{{\rm had}}(E',X)\,\left[1-K(E'/E_{i+1}^{{\rm had}})\right]
\label{source2}\\
W_{d\rightarrow a}^{ij}&=&\int_{\max\left[E_{i-1}^{{\rm had}},E_{\min}^{{\rm had}}
\right]}^{E_{i}^{{\rm had}}}\! dE'\; W_{d\rightarrow a}(E',E_{j}^{{\rm had}})\;
 K(E'/E_{i}^{{\rm had}})\nonumber\\ 
 &  & +\int_{E_{i}^{{\rm had}}}^{\min\left[E_{i+1}^{{\rm had}},E_{{\rm thr}}\right]}
 \! dE'\; W_{d\rightarrow a}(E',E_{j}^{{\rm had}})\,\left[1-K(E'/E_{i+1}^{{\rm had}})\right],
\label{wij2}
\end{eqnarray}
 and the discrete energy loss term via :\[
-\beta_{a}^{{\rm ion}}(E_{i}^{{\rm had}})\:\frac{h_{a}^{i}(X)}
{E_{i}^{\mathrm{had}}-E_{i-1}^{\mathrm{had}}}
+\beta_{a}^{{\rm ion}}(E_{i+1}^{{\rm had}})\:\frac{h_{a}^{i+1}(X)}
{E_{i+1}^{\mathrm{had}}-E_{i}^{\mathrm{had}}}\,.\]
Here the condition of both energy and particle number conservation
gives
\begin{equation}
K(\varepsilon)=\frac{C\:\varepsilon-1}{C-1}\,.
\label{keps}
\end{equation}

The solution of the homogeneous part of Eq. (\ref{sys2-had}) is
\begin{equation}
h_{a}^{i}(X)  = h_{a}^{i}(X_{0})\:
\exp\!\left[-\left(\frac{1-W_{a\rightarrow a}^{ii}}{\lambda_{a}(E_{i}^{{\rm had}})}
+\frac{\beta_{a}^{{\rm ion}}(E_{i}^{{\rm had}})}
{E_{i}^{{\rm had}}-E_{i-1}^{{\rm had}}}\right)(X-X_{0})
-\frac{m_{a}\left|L(X)-L(X_{0})\right|}{c\tau_{a}^{(0)}E_{i}^{{\rm had}}}\right].
\label{hhomo}
\end{equation}

Correspondingly the solution of the full equation can be obtained
in an iterative way 
\begin{eqnarray}
h_{a}^{i}(X) & = & h_{a}^{i}(X_{0})\:\exp\!\left[-\left(\frac{1-W_{a\rightarrow a}^{ii}}{\lambda_{a}(E_{i}^{{\rm had}})}+\frac{\beta_{a}^{{\rm ion}}(E_{i}^{{\rm had}})}{E_{i}^{{\rm had}}-E_{i-1}^{{\rm had}}}\right)(X-X_{0})-\frac{m_{a}\left|L(X)-L(X_{0})\right|}{c\tau_{a}^{(0)}E_{i}^{{\rm had}}}\right]
\nonumber\\
& & +  \int_{X_{0}}^{X}\! dX'\left\{ \sum_{d}\sum_{j=i+1}^{i_{\max}}h_{d}^{j}(X')\,\left[\frac{W_{d\rightarrow a}^{ji}}{\lambda_{d}(E_{j}^{{\rm had}})}+D_{d\rightarrow a}^{ji}\frac{m_{d}/(c\tau_{d}^{(0)})}{E_{j}^{{\rm had}}\rho_{{\rm air}}(X')}\right]\right.
\nonumber \\
 &  & \left.+h_{a}^{i+1}(X')\frac{\beta_{a}^{{\rm ion}}(E_{i+1}^{{\rm had}})}{E_{i+1}^{\mathrm{had}}-E_{i}^{\mathrm{had}}}+S_{ai}^{{\rm had}}(X')\right\} 
\nonumber \\
 &  & \times\exp\!\left[-\left(\frac{1-W_{a\rightarrow a}^{ii}}
 {\lambda_{a}(E_{i}^{{\rm had}})}+\frac{\beta_{a}^{{\rm ion}}(E_{i}^{{\rm had}})}
 {E_{i}^{{\rm had}}-E_{i-1}^{{\rm had}}}\right)(X-X')
 -\frac{m_{a}\left|L(X)-L(X')\right|}{c\tau_{a}^{(0)}E_{i}^{{\rm had}}}\right].
\label{hfull} \end{eqnarray}

Discretizing depth positions as $X_{m}=m\,\Delta\! X$, $m=1,...,m_{\max}$,
$X_{m_{\max}}$ corresponding to the observation level for the given shower
trajectory, the formulas (\ref{hhomo}-\ref{hfull}) can be used to calculate
spectra of different hadrons at all depths $X_{m}$, starting from
the initial condition $h_{a}^{i}(X_{m})$ for $m=1$, calculating
first $h_{a}^{i_{\max}}(X_{m+1})$, using Eq. (\ref{hhomo}), and
$h_{a}^{i_{\max}-1}(X_{m+1})$, $h_{a}^{i_{\max}-2}(X_{m+1})$, ...
, $h_{a}^{1}(X_{m+1})$, using Eq. (\ref{hfull}), etc. for $m=2,...,m_{\max}$.
The depth integral in Eq. (\ref{hfull}) is taken using the Simpson
formula; particle spectra values $h_{a}^{i}(X_{m+1/2})$ at mids of
the bins $X_{m+1/2}=(X_{m}+X_{m+1})/2$ are obtained via a logarithmic
interpolation between previously calculated $h_{a}^{i}(X_{m})$, $h_{a}^{i}(X_{m+1})$.

As for neutral pions, we assume them to decay at the place, calculate
the number of $\pi^{0}$s produced at depth $X_{m}$ as
\begin{eqnarray}
h_{\pi^{0}}^{i}(X_{m})&=&\int_{X_{m-1}}^{X_{m}}\! dX'\sum_{d}\sum_{j=i+1}^{i_{\max}}
\, h_{d}^{j}(X')
\nonumber\\
&  & \times\left[\frac{W_{d\rightarrow\pi^{0}}^{ji}}{\lambda_{d}(E_{j}^{{\rm had}})}
+D_{d\rightarrow\pi^{0}}^{ji}\frac{m_{d}/(c\,\tau_{d}^{(0)})}{E_{j}^{{\rm had}}\;
\rho_{{\rm air}}(X')}\right]+\int_{X_{m-1}}^{X_{m}}\! dX'\: S_{\pi^{0}i}^{{\rm had}}(X')
\label{pi0}
\end{eqnarray}
and add photons resulting from $\pi^{0}$ decay to the e/m source function.

\subsection{Numerical treatment of e/m cascade equations\label{sub:Numerical-em}}

To reduce the problem to the solution of a system of ordinary differential
equations we again, like in case of hadron cascade, discretize particle
energy spectra $l_{a}\!(E,X)$ using an energy grid $E_{i}^{{\rm e/m}}=E_{\min}^{{\rm e/m}}\, C^{i-1}$
($E_{\min}^{{\rm e/m}}=0.1\div1$ MeV), replace the integral over
parent particle energy $E'$ in (\ref{sys1-el}-\ref{sys1-gam}) by
a sum over discrete energies, and discretize the source term $S_{a}^{{\rm e/m}}\!(E,X)$
and the differential energy spectra $W_{d\rightarrow a}(E',E)$ according
to (\ref{source2}-\ref{wij2}) (with $E_{i}^{{\rm had}}$, $E_{\min}^{{\rm had}}$,
$S_{a}^{{\rm had}}\!(E',X)$ being replaced by $E_{i}^{{\rm e/m}}$,
$E_{\min}^{{\rm e/m}}$, $S_{a}^{{\rm e/m}}\!(E',X)$). Then 
Eqs.~(\ref{sys1-el}-\ref{sys1-gam})
are transformed to
\begin{eqnarray}
&  & \frac{dl_{a}^{i}(X)}{dX}=\sum_{d}\sum_{j=i}^{i_{\max}}
\bar{W}_{d\rightarrow a}^{ji}\; l_{d}^{j}(X)+S_{ai}^{{\rm e/m}}\!(X),
\label{sys2-em}
\end{eqnarray}
where we replaced the continuous energy loss term
$\frac{\partial}{\partial E}\left( \beta_{e^{\pm}}^{{\rm ion}}(E)
 l_{e^{\pm}}(E,X)\right)$ 
by
\[
-\beta_{e^{\pm}}^{{\rm ion}}\!(E_{i}^{{\rm e/m}})\;
\frac{l_{e^{\pm}}^{i}\!(X)}{E_{i}^{{\rm e/m}}-E_{i-1}^{{\rm e/m}}}
+\beta_{e^{\pm}}^{{\rm ion}}\!(E_{i+1}^{{\rm e/m}})\;
\frac{l_{e^{\pm}}^{i+1}\!(X)}{E_{i+1}^{{\rm e/m}}-E_{i}^{{\rm e/m}}}
\]
 and included the two terms, together with the interaction cross sections,
into the discrete particle production spectra $\bar{W}_{d\rightarrow a}^{ji}$,
defined as follows
\begin{eqnarray}
 &  & \bar{W}_{e^{\pm}\rightarrow e^{\pm}}^{ii}=W_{e^{\pm}\rightarrow e^{\pm}}^{ii}-\sigma_{e^{\pm}}\!(E_{i}^{{\rm e/m}})-\frac{\beta_{e^{\pm}}^{{\rm ion}}\!(E_{i}^{{\rm e/m}})}{(C-1)E_{i}^{{\rm e/m}}}
\label{wii-ep}\\
 &  & \bar{W}_{e^{\pm}\rightarrow e^{\pm}}^{i,i-1}=W_{e^{\pm}\rightarrow e^{\pm}}^{i,i-1}+\frac{\beta_{e^{\pm}}^{{\rm ion}}\!(E_{i}^{{\rm e/m}})}{(C-1)E_{i}^{{\rm e/m}}}
\label{wii1-ep}\\
 &  & \bar{W}_{\gamma\rightarrow\gamma}^{ii}=W_{\gamma\rightarrow\gamma}^{ii}-\sigma_{\gamma}\!(E_{i}^{{\rm e/m}}),\label{wiigg}
\end{eqnarray}
and $\bar{W}_{d\rightarrow a}^{ij}=W_{d\rightarrow a}^{ij}$ for all
other combinations of $i$, $j$, $a$, $d$. 

It is worth verifying that all elements of the matrixes $\bar{W}_{d\rightarrow a}^{ij}$
are free of singularities. Indeed, singular terms appear in 
$\sigma_{e^{\pm}}^{{\rm brems}}\!(E_{i}^{{\rm e/m}})=
\int_{0}^{E_{i}^{{\rm e/m}}}\! dE'\; W_{e^{-}\rightarrow e^{-}}^{{\rm brems}}\!(E_{i}^{{\rm e/m}},E')$
and in $W_{e^{\pm}\rightarrow e^{\pm}}^{ii}$ (defined by Eq. (\ref{wij2}))
due to the characteristic $1/(E-E')$ dependence of $W_{e^{-}\rightarrow e^{-}}^{{\rm brems}}\!(E,E')$.
Nevertheless, in $\bar{W}_{e^{\pm}\rightarrow e^{\pm}}^{ii}$ defined
by Eq. (\ref{wii-ep}) the corresponding singularities are canceled
against each other
\begin{eqnarray}
\bar{W}_{e^{\pm}\rightarrow e^{\pm}}^{({\rm brems})ii}&=&
W_{e^{\pm}\rightarrow e^{\pm}}^{({\rm brems})ii}-\sigma_{e^{\pm}}^{{\rm brems}}\!(E_{i}^{{\rm e/m}})\\
 &=&\int_{E_{i-1}^{{\rm e/m}}}^{E_{i}^{{\rm e/m}}}\! dE'\;
  W_{e^{-}\rightarrow e^{-}}^{{\rm brems}}\!(E_{i}^{{\rm e/m}},E')\;
  \frac{C\: E'/E_{i}^{{\rm e/m}}-1}{C-1}\nonumber\\
 &  & -\int_{0}^{E_{i}^{{\rm e/m}}}\! dE'\;
  W_{e^{-}\rightarrow e^{-}}^{{\rm brems}}\!(E_{i}^{{\rm e/m}},E')\\
 &=& -\frac{C}{E_{i}^{{\rm e/m}}\:(C-1)}
 \int_{E_{i-1}^{{\rm e/m}}}^{E_{i}^{{\rm e/m}}}\! dE'\;
 (E_{i}^{{\rm e/m}}-E')\; W_{e^{-}\rightarrow e^{-}}^{{\rm brems}}\!(E_{i}^{{\rm e/m}},E')\nonumber\\
 &  & -\int_{0}^{E_{i-1}^{{\rm e/m}}}\! dE'\; W_{e^{-}\rightarrow e^{-}}^{{\rm brems}}\!(E_{i}^{{\rm e/m}},E'),
\end{eqnarray}
where the last two integrals are finite. Similarly one can check the
finiteness of $\bar{W}_{e^{\pm}\rightarrow e^{\pm}}^{i,i-1}$.

To solve the system (\ref{sys2-em}), we first find the solution of
the homogeneous equation system
\begin{eqnarray}
 &  & \frac{dl_{a}^{i}(X)}{dX}=\sum_{d}\bar{W}_{d\rightarrow a}^{ii}\;
  l_{d}^{i}(X),
\label{sys-emhomo}
\end{eqnarray}
which is given as
\begin{eqnarray}
 &  & l_{a}^{i}(X)=\sum_{I=1}^{3}D_{aI}^{i}(X_{0})\; e^{\Lambda_{I}^{i}\:(X-X_{0})}
\label{lept-homo}
\end{eqnarray}
Here $\Lambda_{I}^{i}$ and $D_{aI}^{i}(X_{0})$ are correspondingly
the eigenvalues and eigenvectors of the system of linear algebraic
equations obtained by inserting (\ref{lept-homo}) into (\ref{sys-emhomo}):\[
\Lambda^{i}\; D_{a}^{i}(X_{0})-\sum_{d}\bar{W}_{d\rightarrow a}^{ii}\;
 D_{d}^{i}(X_{0})=0,\]
with the normalization fixed by the initial conditions at $X=X_{0}$: 
$\sum_{I=1}^{3}D_{aI}^{i}(X_{0})=l_{a}^{i}(X_{0})$.

Then the solution of the full equation system (\ref{sys2-em}) may
be given in a recursive form
\begin{eqnarray}
l_{a}^{i}(X)&=&\sum_{I=1}^{3}D_{aI}^{i}(X_{0})\:
\left[e^{\Lambda_{I}^{i}\:(X-X_{0})}+\sum_{d}F_{Id}^{i}(X_{0})
\int_{X_{0}}^{X}\! dX'\right.
\nonumber\\
 &  & \left.\times\: e^{\Lambda_{I}^{i}\:(X-X')}\:\left[S_{di}^{{\rm e/m}}\!(X')
 +\sum_{j=i+1}^{i_{\max}}\sum_{g}\bar{W}_{g\rightarrow d}^{ji}\;
  l_{g}^{j}(X')\right]\right],
\label{lept-full}
\end{eqnarray}
where the matrix $F_{Id}^{i}$ is inverse with respect to $D_{aI}^{i}$
\begin{equation}
\sum_{I}D_{aI}^{i}(X_{0})\; F_{Id}^{i}(X_{0})=\delta_{a}^{d}\,.
\label{inver}
\end{equation}

Equations (\ref{lept-homo}-\ref{lept-full}) can be used to calculate
the discrete spectra of e/m particles $l_{a}^{i}(X_{m})$ at all
depths $X_{m}$ in the same way as in case of the hadronic cascade.

\paragraph*{Acknowledgments}
The authors thank Hans-Joachim Drescher, Michael Unger, and 
Ralf Ulrich for fruitful discussions. N.N.K. acknowledges the 
financial support of the Russian Foundation for Basic Research (RFBR,
grant 05-02-16401).
The work of one of the authors, S.O., has been supported in part by 
the German Ministry for Education
and Research (BMBF, Grant 05 CU1VK1/9).


\end{document}